\documentclass[aps,pre,twocolumn,groupedaddress,floatfix,longbibliography,superscriptaddress]{revtex4-1}

\usepackage{amsmath}
\usepackage{amssymb}
\usepackage{amsfonts}
\usepackage{graphicx}
\usepackage{bbold}
\usepackage{bm}
\usepackage{bm}
\usepackage{cancel}
\usepackage{times,float}
\usepackage{graphicx}
\usepackage[usenames,dvipsnames,svgnames]{xcolor}
\usepackage{hyperref}
\hypersetup{colorlinks=true, linkcolor=Blue, citecolor=Blue,urlcolor=Blue}
\usepackage{multirow}
\usepackage{ulem}
\usepackage{color,epsfig}
\usepackage{bm}
\usepackage{slashed}
\usepackage{hyperref}
\usepackage{subfigure}
\usepackage{feynmf}
\usepackage{mathrsfs}

\newcommand{\bea}{\begin{eqnarray}}
\newcommand{\eea}{\end{eqnarray}}
\newcommand{\be}{\begin{equation}}
\newcommand{\ee}{\end{equation}}

\newcommand{\sgn}{{\rm sign}}

\newcommand{\nn}{\nonumber}
\newcommand{\ii}{\mathrm{i}}

\newcommand{\qB}{|q_f B|}

\newcommand{\kt}{\mathbf{k}_\perp}
\newcommand{\pt}{\mathbf{p}_\perp}

\newcommand{\kp}{k_\parallel}
\newcommand{\gt}{g_\perp}
\newcommand{\gp}{g_\parallel}
\newcommand{\gn}{\gamma^\nu}
\newcommand{\gm}{\gamma^\mu}

\newcommand{\pp}{p_\parallel}

\newcommand{\lp}{\ell_\parallel}
\newcommand{\Tr}[1]{\text{Tr}\left\{{#1}\right\}}
\newcommand{\Op}[1]{\mathcal{O}^{(#1)}}
\newcommand{\T}[1]{t_{#1}^{\mu\nu}}

\begin{document}

\title{Exploring magnetic fluctuations effects in QED gauge fields: implications for mass generation}
\author{Jorge David Casta\~no-Yepes}
\email{jcastano@uc.cl}
\affiliation{Facultad de F\'isica, Pontificia Universidad Cat\'olica de Chile, Vicu\~{n}a Mackenna 4860, Santiago, Chile}
\author{Enrique Mu\~noz}
\email{ejmunozt@uc.cl}
\affiliation{Facultad de F\'isica, Pontificia Universidad Cat\'olica de Chile, Vicu\~{n}a Mackenna 4860, Santiago, Chile}
\affiliation{Center for Nanotechnology and Advanced Materials CIEN-UC, Avenida Vicuña Mackenna 4860, Santiago, Chile}

\begin{abstract}
In this work, we calculate the one-loop contribution to the polarization tensor for photons (and gluons) in the presence of a classical background magnetic field with white-noise stochastic fluctuations. The magnetic field fluctuations are incorporated into the fermion propagator in a quasi-particle picture, which we developed in previous works using the {\it replica trick}. By focusing on the strong-field limit, here we explicitly calculate the polarization tensor. Our results reveal that it does not satisfy the transversality conditions outlined by the Ward identity, thus breaking the $U(1)$ symmetry. As a consequence, in the limit of vanishing photon four-momenta, the tensor coefficients indicate the emergence of an effective magnetic mass induced on photons (and gluons) by these stochastic fluctuations, leading to the interpretation of a dispersive medium with a noise-dependent index of refraction.
\end{abstract}

\maketitle 
\section{Introduction}
Understanding the properties of photons and gluons in thermal and magnetized media is essential for a proper interpretation of the observables arising from current high-energy experiments~\cite{David_2020,PhysRevC.91.064904}. Recent studies on photon production in heavy-ion collisions (HIC) have revealed that photons originating from such scenarios exhibit an elliptic flow coefficient, denoted as $v_2$, of similar magnitude to that measured in hadrons~\cite{PhysRevLett.109.122302,2019308,PhysRevC.94.064901}. Various sources of photons have been proposed to characterize the observed spectrum. A significant photon yield is generated during the equilibrium stages of HIC, which is utilized to estimate the temperature of the colliding system~\cite{GHIGLIERI2014326}. On the other hand, direct photons are believed to be produced during the hadronization stages of HIC, where much of the $v_2$ is generated. Furthermore, prompt photons are identified as part of the low $p_\text{T}$ spectra~\cite{PAQUET2016409}. Despite the aforementioned identified sources of photons, there remains a discrepancy between the theoretical models developed to describe the photon spectra and the corresponding experimental measurements. In particular, an excess of low-$p_\text{T}$ photons is obtained when comparing theory and experimental data~\cite{PhysRevC.93.044906}. 

In an effort to provide a more comprehensive description of the experimental data, recent studies have suggested that the production of prompt photons may be influenced by the intense magnetic fields generated during the initial stages of the collision~\cite{PhysRevC.106.064905, Ayala2020, Ayala:2017vex,JIA2023138239}. These investigations propose that the background magnetic field induces gluon fusion and splitting processes for photon generation due to the high-density gluon occupation, referred to as the Color Glass Condensate (CGC)~\cite{LAPPI2006200,MCLERRAN201471, Harland-Lang2015, PhysRevC.106.034904}. Although this hypothesis leads to an improved description of the elliptic flow and yield, the kinematic restrictions arising from the vanishing mass of photons and gluons reduce the available phase space for the number of photons~\cite{PhysRevLett.25.1061}. Therefore, any modification in the dispersion relations for gluons and photons induced by the medium may open up a richer physical scenario. For instance, it is well-known that a thermalized medium can induce an effective photon/gluon mass~\cite{PhysRevD.28.908, lebellac}. Similar effects may in principle arise from a magnetized medium, described by a noisy background magnetic field, and this is the main focus of the present work. 

Very intense magnetic fields are generated in semi-central HIC by the presence of spectator particles, but they rapidly decay~\cite{PhysRevC.83.054911,PhysRevLett.127.052302}. This leads to an incomplete electromagnetic response in the effective medium formed at later times, such as the Quark-Gluon Plasma (QGP)~\cite{PhysRevC.105.L041901}. Consequently, the magnetic field is found to be particularly intense during the pre-equilibrium stage. Nevertheless, from a theoretical perspective, the screening effects from magnetized media do not modify the dispersion relation when a constant magnetic field is taken into account~\cite{Ayala2021}.  

The absence of an induced photon/gluon mass in a uniformly magnetized medium arises from symmetry considerations. In the context of the one-loop polarization tensor approximation, the $U(1)$ symmetry remains intact in a constant background magnetic field, as the Ward-Takahashi identity is still satisfied under such condition~\cite{Ayala_Pol_020}. As a consequence, the inverse propagator for gauge fields continues to exhibit a pole at $q^2=0$, thus implying the absence of an induced magnetic mass. Therefore, a physical condition that breaks such a symmetry may lead to the generation of a magnetic gluon/photon mass.

In two of our recent works~\cite{PhysRevD.107.096014, PhysRevD.108.116013}, we investigated the implications of classical background magnetic field, possessing stochastic fluctuations, on the properties of a QED medium. We model this scenario by assuming that the classical background magnetic field arises from a classical gauge field $A_\text{BG}^{i}(x) + \delta A_\text{BG}^{i}(x)$ with white-noise correlated stochastic fluctuations $\delta A_\text{BG}^{i}(x)$, as described by the statistical properties~\cite{PhysRevD.107.096014, PhysRevD.108.116013}
\bea
\langle \delta A_\text{BG}^{i}(x)\rangle &=& 0\nn\\
\langle \delta A_\text{BG}^{i}(x) \delta A_\text{BG}^{j}(y)\rangle &=& \Delta_B \delta_{ij}\delta(x-y).
\eea

By applying the so-called {\it replica trick}~\cite{mezard1991replica}, we derived an effective interaction term for QED fermions in the presence of such a noisy magnetic field. This approximation leads, at the perturbative level, to a renormalization of the fermion propagator that now represents quasiparticles propagating in a dispersive medium~\cite{PhysRevD.107.096014}. On the other hand, when we apply our analysis at the mean field level, the emergence of vector currents is predicted~\cite{PhysRevD.108.116013}. Both perspectives point towards violations of $U(1)$ symmetry due to the presence of stochastic fluctuations in the background magnetic field.

In this work, we apply our results obtained in Ref.~\cite{PhysRevD.107.096014} to calculate the one-loop polarization tensor for photons and gluons, and we find that it is explicitly non-transverse in the sense of the Ward-Takahashi identity. This effect arises from to the breaking of $U(1)$ symmetry due to the incoherent, stochastic nature of the classical background gauge field $A_\text{BG}^{\mu}(x) + \delta A_\text{BG}^{\mu}(x)$, thus resulting in the generation of a dynamical magnetic mass for the quantum gauge fields (photons and gluons). 
\section{The photon polarization tensor at strong magnetic field}
\begin{figure}[h!]
    \centering
    \includegraphics[scale=0.27]{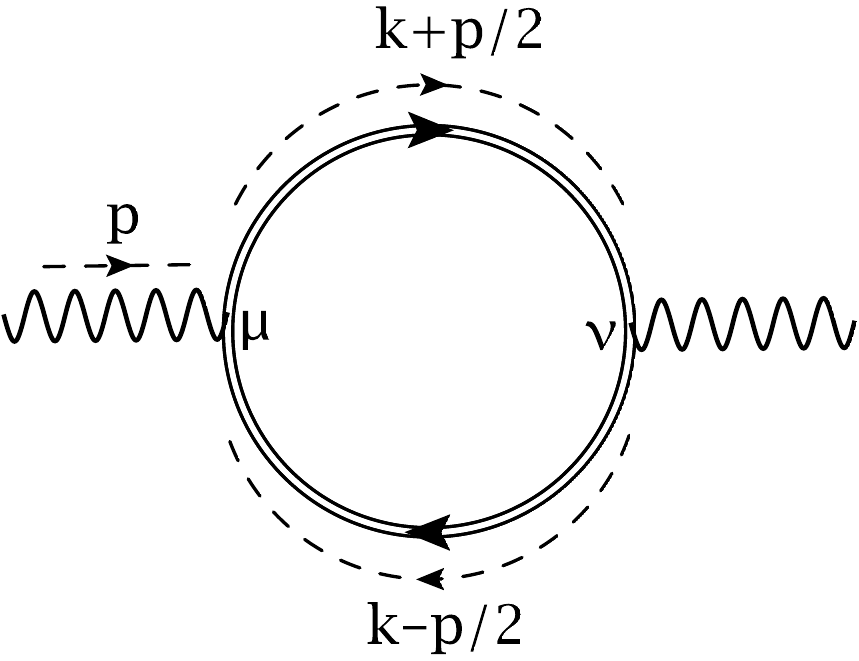}\hspace{0.2cm}\includegraphics[scale=0.27]{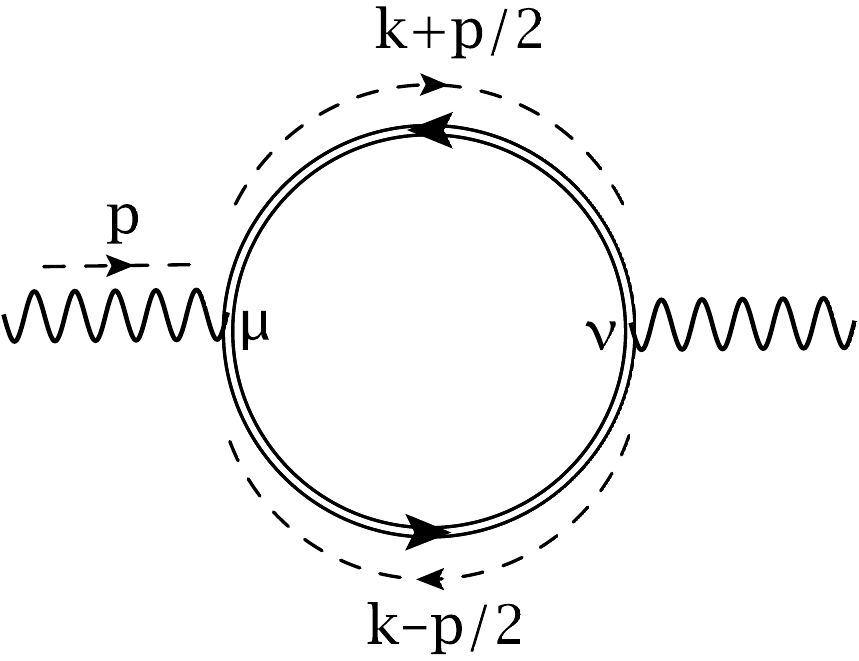}
    \caption{Feynmann diagrams that contribute to the one-loop photon polarization tensor. The arrows in the propagators represent the direction of the flow of charge, whereas the dashed arrows represent the momentum flux.}
    \label{fig:polarization}
\end{figure}

Our starting point is the one-loop contribution to the photon/gluon polarization tensor, which is depicted in Fig.~\ref{fig:polarization} and is given by
{\small
\bea
\ii\Pi_\Delta^{\mu\nu}&=&-\frac{1}{2}\int\frac{d^4k}{(2\pi)^4}\text{Tr}\Big\{\ii q_f\gn\ii S_\Delta^{(-)}\left(k\right)\ii q_f\gm\ii S_\Delta^{(-)}(k-p)\Big\}\nn\\
&-&\frac{1}{2}\int\frac{d^4k}{(2\pi)^4}\text{Tr}\Big\{\ii q_f\gn\ii S_\Delta^{(+)}(-k+p)\ii q_f\gm\ii S_\Delta^{(+)}(-k)\Big\},
\label{eq:PloTenDef}\nn\\
\eea
}
where $q_f$ is the electric charge of the fermion in the loop, and the antiparticle or charge conjugated contribution has been taken into account.

Note that the only difference between the photon and gluon polarization tensors arises from the trace over color $SU(3)$ space, i.e.,
\bea
\ii\Pi_\text{gluon}^{\mu\nu}=\Tr{t_at_b}\ii\Pi_\text{photon}^{\mu\nu}=\frac{1}{2}\delta_{ab}\,\ii\Pi_\text{photon}^{\mu\nu},
\eea
where $t_{a,b}$ are the generators of the color group in the fundamental representation. 

\begin{figure}
    \centering
    \includegraphics[scale=0.5]{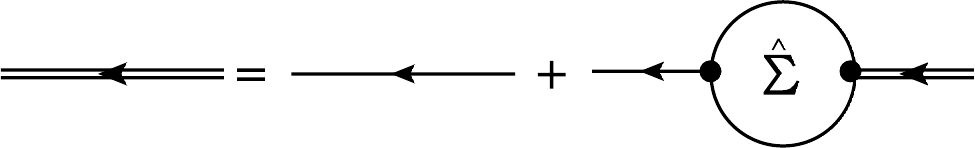}
    \caption{Dyson equation for the "dressed" propagator (double-line), in terms of the free propagator (single-line) and the self-energy $\Sigma$.}
    \label{fig:DiagramSelfEnergy2}
\end{figure}

\begin{figure}[h!]
    \centering
    \includegraphics[scale=0.4]{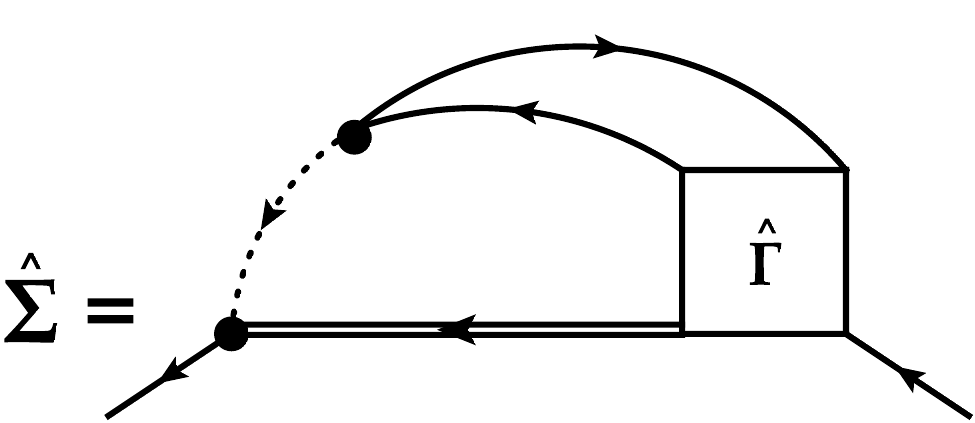}
    \caption{Skeleton diagram representing the self-energy for the effective interacting theory. The dashed line is the disorder-induced interaction $\Delta_B$, while the box $\hat{\Gamma}$ represents the 4-point vertex function.}
    \label{fig:DiagramSelfEnergy1}
\end{figure}

To analytically compute Eq.~\eqref{eq:PloTenDef}, we will employ the renormalized fermion propagator in the presence of static (quenched) white noise spatial fluctuations, focusing on the regime of a strong external magnetic field~\cite{PhysRevD.107.096014,PhysRevD.108.116013}. Specifically, the effective fermion-fermion interaction arises as a result of averaging over the background magnetic noise, where we include the magnetic noise-induced interaction effects by dressing the Schwinger propagator with a self-energy, as shown diagrammatically in the Dyson equation depicted in Fig.~\ref{fig:DiagramSelfEnergy2}. We remark that for this theory, the skeleton diagram for the self-energy is represented in Fig.~\ref{fig:DiagramSelfEnergy1}, and the dressed propagator is provided by

\begin{figure}[h!]
    \centering
    \includegraphics[scale=0.5]{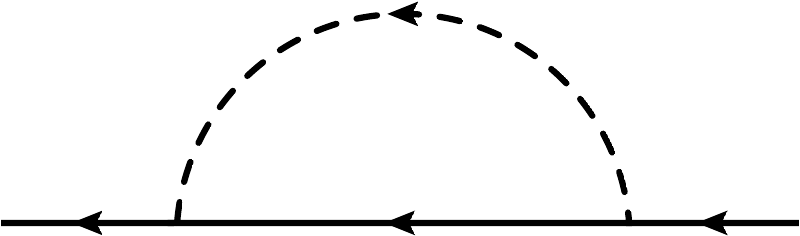}
    \caption{Self-energy diagram at first order in $\Delta = q_f^2\Delta_B$.}
    \label{fig:selfenergydiagram}
\end{figure}
\bea
\ii S_\Delta^{(\pm)}(p)&=&C(p)\left(m+\gamma_0 p^0+\frac{\gamma_3 p^3}{z(p)}\right)\mathcal{P}^{(\pm)}(p),\nn\\
\label{eq:Prop_Delta}
\eea
where $m$ is the fermion mass and
\begin{subequations}
\bea
C(p)&=&\ii\frac{z(p)\,e^{-\mathbf{p}_\perp^2/|q_f B|}}{p_\parallel^2-z^2(p)m^2},
\eea
\bea
z(p)=1+\frac{3}{4}\frac{\Delta|q_f B|e^{-\mathbf{p}_\perp^2/|q_f B|}}{\pi\sqrt{p_0^2-m^2}}.
\eea
\bea
z_3(p)=\frac{1}{z(p)}\left(1+\frac{\Delta|q_f B|e^{-\mathbf{p}_\perp^2/|q_f B|}}{4\pi\sqrt{p_0^2-m^2}}\right),
\eea

\bea
\mathcal{P}^{(\pm)}(p)=\frac{1}{2}\left[1\pm\sgn(q_f B)\ii z_3(p)\gamma^1\gamma^2\right].
\eea
\end{subequations}

As explained in Ref.~\cite{PhysRevD.107.096014}, the fermion propagator self-energy was computed to order $\Delta \equiv q_f^2 \Delta_B$. Therefore, to maintain consistency with this level of approximation, we expand Eq.~\eqref{eq:Prop_Delta} as follows:
\bea
\ii S_\Delta^{(\pm)}(p)&=&\ii S_0^{(\pm)}(p)+\ii\Delta\left(\frac{\qB}{2\pi}\right)\Big[\Theta_1(p)(\slashed{p}_\parallel+m)\mathcal{O}^{(\pm)}\nn\\
&-&\Theta_2(p)\gamma^3\mathcal{O}^{(\pm)}\pm\Theta_3(p)\ii\gamma^1\gamma^2(\slashed{p}_\parallel+m)\Big]\nn\\
&+&\text{O}(\Delta^2),
\eea
where
\bea
\ii S_0^{(\pm)}(p)&=&2\ii\frac{e^{-\pt^2/|q_f B|}}{\pp^2-m^2}(\slashed{p}_\parallel+m)\mathcal{O}^{(\pm)}
\eea
is the fermion propagator in the presence of an intense magnetic field, the spin-projection operator is given by
\bea
\mathcal{O}^{(\pm)}&=&\frac{1}{2}\left[1\pm\sgn(q_f B)\ii\gamma^1\gamma^2\right],
\eea
and we defined the functions
\begin{subequations}
\bea
\Theta_1(p)&\equiv&\frac{3(\pp^2+m^2)e^{-2\pt^2/|q_f B|}}{(\pp^2-m^2)^2\sqrt{p_0^2-m^2}} ,
\eea
\bea
\Theta_2(p)&\equiv&\frac{3p_3e^{-2\pt^2/|q_f B|}}{(\pp^2-m^2)\sqrt{p_0^2-m^2}},
\eea
\bea
\Theta_3(p)&\equiv&\frac{e^{-2\pt^2/|q_f B|}}{(\pp^2-m^2)\sqrt{p_0^2-m^2}}.
\eea
\end{subequations}

Here, we separated the parallel ($\parallel$) from the perpendicular ($\perp$) Minkowski subspaces, as defined by their relative direction with respect to the background external magnetic field, by splitting the metric tensor as
\begin{subequations}
    \bea
g^{\mu\nu}=\gp^{\mu\nu}+\gt^{\mu\nu},
\eea
where
\bea
\gp^{\mu\nu}&=&\text{diag}(1,0,0,-1)\nn\\
\gt^{\mu\nu}&=&\text{diag}(0,-1,-1,0).
\eea
\end{subequations}

The latter implies that for any four-vector
\begin{subequations}
    \bea
    p^\mu&=&\pp^\mu+p_\perp^\mu,
\eea
we get
\bea
p^2=\pp^2-\pt^2,
\eea
with
\bea
\pp^2&=&p_0^2-p_3^2\nn\\
\pt^2&=&p_1^2+p_2^2.
\eea
\end{subequations}


Hence, Eq.~\eqref{eq:PloTenDef} takes the form:
\bea
\ii\Pi_\Delta^{\mu\nu}&=&\ii\Pi_0^{\mu\nu}+\ii\frac{q_f^2\qB\Delta}{4\pi}\sum_{i=1}^{3}T_i^{\mu\nu},
\eea
where $\ii\Pi_0^{\mu\nu}$ is the one-loop polarization tensor in the strong field limit and in the absence of fluctuations~\cite{PhysRevD.83.111501}:
\bea
\ii\Pi_0^{\mu\nu}=\frac{\ii q_f^2\qB}{4\pi^2}e^{-\pt^2/2\qB}\pp^2\mathcal{I}_0(\pp^2)\left(\gp^{\mu\nu}-\frac{\pp^\mu\pp^\nu}{\pp^2}\right)
\eea
so that  (see Appendix~\ref{sec:PropagatorAtorderDelta})
\bea
\mathcal{I}_0(\pp^2)&=&\int_0^1dx\frac{x(1-x)}{x(1-x)\pp^2-m^2}\\
&=& \frac{1}{p_{\parallel}^2}\left[ 1 + \frac{2 m^2/p_{\parallel}^2}{\sqrt{1 - \frac{4 m^2}{p_{\parallel}^2}}}\log\left[ \frac{1 + \sqrt{1 - \frac{4 m^2}{p_{\parallel}^2}}}{1 - \sqrt{1 - \frac{4 m^2}{p_{\parallel}^2}}} \right]  \right],\nn
\eea
and the tensors $T_i^{\mu\nu}$ are given by:

\begin{widetext}
\begin{subequations}
\bea   
T_1^{\mu\nu} &\equiv& 16\ii\int\frac{d^4k}{(2\pi)^4}\frac{e^{-\kt^2/\qB}}{\kp^2-m^2}\Theta_1(k-p)\Bigg[(m^2+\kp\cdot(\pp-\kp))(g_\parallel^{\mu\nu}-g_\perp^{\mu\nu})+(\kp^\mu - \pp^\mu)\kp^\nu+ \kp^\mu(\kp^\nu - \pp^\nu)\Bigg],
\eea
\bea
T_2^{\mu\nu}&\equiv&16\ii\int\frac{d^4k}{(2\pi)^4}\frac{ e^{-\kt^2/\qB}}{\kp^2-m^2}\Theta_2(k-p)\left(k^3g_\parallel^{\mu\nu}+\kp^\mu\delta_3^\nu+\kp^\nu\delta_3^\mu\right)
\eea
    and

\bea
T_3^{\mu\nu} &\equiv& 16\ii\int\frac{d^4k}{(2\pi)^4}\frac{e^{-\kt^2/\qB}}{\kp^2-m^2}\Theta_3(k-p)\Bigg[(m^2+\kp\cdot(\pp-\kp))(g_\parallel^{\mu\nu}-g_\perp^{\mu\nu})+(\kp^\mu - \pp^\mu)\kp^\nu+ \kp^\mu(\kp^\nu - \pp^\nu)\Bigg].
\eea
\label{T1T2T3}
\end{subequations}
\end{widetext}

Further details are presented in the Appendix~\ref{sec:momentumintegrals}.

\section{The gauge field mass generation by magnetic fluctuations }
To ascertain the role of the magnetic field fluctuations on the possible generation of mass in the gauge fields, we identify the poles of the propagator. From the Dyson equation, its inverse is

\begin{equation}
\left[D^{\mu\nu}(p)\right]^{-1} = \left[D^{\mu\nu}_0(p)\right]^{-1} - \ii\Pi^{\mu\nu}(p),
\end{equation}

where

\begin{equation}
D^{\mu\nu}_0(p) = \frac{-\ii\,g^{\mu\nu}}{(p^2 + \ii\epsilon)}
\end{equation}
is the ``free" photon propagator, in the Feynman gauge. Furthermore, we shall approximate the polarization tensor up to one-loop, by applying the results obtained in our previous calculations.

Following the approach outlined in Ref.~\cite{lebellac}, the poles associated with the dynamic mass emerge as the coefficients of $g^{\mu\nu}_{\parallel}$ and $g^{\mu\nu}_{\perp}$, respectively, when the limits $p_0\to0$ and $\mathbf{p}\to0$ in $\ii\Pi^{\mu\nu}(p)$ are considered. Calculating those limits in Eqs.~\eqref{T1T2T3}, we obtain
\begin{subequations}
\bea
\lim_{\substack{p_0\to0\\ \mathbf{p}\to0}}T_1^{\mu\nu}&=&\frac{4\ii\qB}{\pi m}\int_{\mathbb{R}^2}\frac{dydx}{(2\pi)^2}\frac{(x^2+y^2-1)\gp^{\mu\nu}}{(x^2+y^2+1)^3\sqrt{y^2+1}}\nn\\
&=& -\frac{\ii \qB}{32\pi m}\gp^{\mu\nu},
\eea
\bea
\lim_{\substack{p_0\to0\\ \mathbf{p}\to0}}T_2^{\mu\nu}&=&-\frac{4\ii\qB}{\pi m}\int_{\mathbb{R}^2}\frac{dydx}{(2\pi)^2}\frac{x^2\left(\gp^{\mu\nu}+2\delta_3^\mu\delta_3^\nu\right)}{(x^2+y^2+1)^2\sqrt{y^2+1}}\nn\\
&=& -\frac{\ii \qB}{2\pi m}\left(\gp^{\mu\nu}+2\delta_3^\mu\delta_3^\nu\right),
\eea 
and
\bea
\lim_{\substack{p_0\to0\\ \mathbf{p}\to0}}T_3^{\mu\nu}&=&\frac{4\ii\qB}{3\pi m}\int_{\mathbb{R}^2}\frac{dydx}{(2\pi)^2}\frac{1}{(x^2+y^2+1)\sqrt{y^2+1}}\nn\\
&&\quad\quad\quad\quad\times\left[\frac{1}{(x^2+y^2+1)^2}\gp^{\mu\nu}-\gt^{\mu\nu}\right]\nn\\
&=& \frac{\ii\qB}{3\pi m}\left[\frac{1}{4}\gp^{\mu\nu}-\gt^{\mu\nu}\right],
\eea
\end{subequations}
where the integrals are computed as explained in detail in Appendix~\ref{sec:momentumintegrals}. Then, after integration and by using the fact that
\bea
\lim_{\substack{p_0\to0\\ \mathbf{p}\to0}}\ii\Pi^{\mu\nu}_0=0,
\eea
we can conclude
\bea
&&\lim_{\substack{p_0\to0\\ \mathbf{p}\to0}}\ii\Pi^{\mu\nu}=\frac{\alpha_\text{em}m^2\mathcal{B}^2}{\pi}\widetilde{\Delta}\left(\frac{43}{96}\gp^{\mu\nu}+\frac{1}{3}\gt^{\mu\nu}+\delta_3^\mu\delta_3^\nu\right),\nn\\
\eea
where
\bea
\alpha_\text{em}\equiv q_f^2/4\pi,~\mathcal{B}\equiv\frac{\qB}{m^2},~\widetilde{\Delta}\equiv m\Delta.
\eea

When these results are substituted into the Dyson equation, using $g^{\mu\nu} = g_{\parallel}^{\mu\nu} + g_{\perp}^{\mu\nu}$, we have that at low-energies the inverse photon propagator is given by 
\bea
\left[D^{\mu\nu}(p)\right]^{-1} &=& \ii g^{\mu\nu}_{\parallel}\left( p^{2} + \ii M_{\parallel}^2 + \ii\epsilon \right)\\ 
&+&  \ii g^{\mu\nu}_{\perp}\left( p^{2} + \ii M_{\perp}^2 + \ii\epsilon \right) + 3 \ii M_{\perp}^2\,\ii \delta_3^\mu\delta_3^\nu\nn + \ldots,
\eea
or inverting this relation,
\bea
D^{\mu\nu}(p) &=& \frac{-\ii g^{\mu\nu}_{\parallel}}{p^{2} + \ii M_{\parallel}^2 + \ii\epsilon} + \frac{-\ii g^{\mu\nu}_{\perp}}{p^{2} + \ii M_{\perp}^2 + \ii\epsilon}\\
&-& \frac{3 M_{\perp}^{2}\delta_3^\mu\delta_3^\nu}{\left( p^2 + \ii M_{\parallel}^2 + \ii\epsilon \right)\left(p^2 + \ii(M_{\parallel}^2 - 3 M_{\perp}^2) + \ii\epsilon  \right)}\nn
\eea
where we defined the magnetic effective masses in both parallel and transverse projections by the coefficients
%
\bea
M_\parallel^2&\equiv&\frac{43\alpha_\text{em}\mathcal{B}^2\widetilde{\Delta}}{96\pi}m^2,\nn\\
M_\perp^2&\equiv&\frac{\alpha_\text{em}\mathcal{B}^2\widetilde{\Delta}}{3\pi}m^2.
\eea

We remark the physical interpretation of those effective masses by comparing with the poles of the photon propagator along each polarization, since 
\be
p^2 + \ii M_{\perp,\parallel}^2 = p_0^2 - \mathbf{p}^2 + \ii M_{\perp,\parallel}^2
\ee
indicates an effective photon dispersion relation for each polarization direction
\be
\omega_{\perp,\parallel}(\mathbf{p}) = \sqrt{\mathbf{p}^2 - \ii M_{\perp,\parallel}^2 },
\ee
where the corresponding effective mass is complex
\be
m_{\perp,\parallel} \equiv \left(-\ii M_{\perp,\parallel}^2  \right)^{1/2} = \frac{1 - \ii}{\sqrt{2}}M_{\perp,\parallel}.
\ee

While the real part represents, as usual, a damping effect, the imaginary part generates an oscillatory component. This picture is consistent with our interpretation of the presence of random magnetic fluctuations as generating an effective dispersive medium both for fermions and photons (or gluons) as well.

\section{Summary and conclusions }
In this study, we computed the one-loop polarization tensor for photons (and gluons) propagating in a medium subjected to a strong magnetic field with white-noise fluctuations. To achieve this, we utilized the fermion propagator developed in our previous work~\cite{PhysRevD.108.116013}, which is obtained through the application of the {\it replica trick} to average the fluctuations over the QED Lagrangian. This approach ensured the consistency of our calculations and allowed us to maintain perturbative accuracy up to order $\text{O}(\Delta)$. Our findings revealed that this tensor does not exhibit transversality in the Ward-Takahashi sense, resulting in the breaking of the system's $U(1)$ symmetry. Furthermore, by following the standard procedure involving the poles of the inverse gauge field propagator, we identified the emergence of magnetic masses generated solely by the fluctuations. Notably, these masses were observed to be distinct in both parallel and perpendicular spatial dimensions, indicating the presence of birefringence effects resulting from the violation of Lorentz symmetry.

\acknowledgements{J.D.C.-Y. and E.M. acknowledge financial support from ANID PIA Anillo ACT/192023. E.M. also acknowledges financial support from Fondecyt 1230440. J.D.C.-Y. also acknowledges financial support from Fondecyt 3220087.

%

\newpage
\appendix
\vspace{5cm}

\begin{widetext}
\section{Propagator at order and polarization tensor $\Delta$}\label{sec:PropagatorAtorderDelta}
We expand $C(p)$, $z_3(p)$, and $1/z(p)$ up to order $\Delta$ as follows:
\bea
C(p)&=&\ii\frac{e^{-\pt^2/\qB}}{\pp^2-m^2}\nn\\
&+&\ii\Delta\left(\frac{3\qB}{4\pi}\right)\frac{(\pp^2+m^2)e^{-2\pt^2/\qB}}{(\pp^2-m^2)^2\sqrt{p_0^2-m^2}}+\text{O}(\Delta^2),\nn\\
\eea
\bea
z_3(p)&=&1-\Delta\left(\frac{\qB}{2\pi}\right)\frac{e^{-\pt^2/\qB}}{\sqrt{p_0^2-m^2}}+\text{O}(\Delta^2)
\eea
\bea
\frac{1}{z(p)}=1-\Delta \left(\frac{3\qB}{4\pi}\right)\frac{e^{-\pt^2/\qB}}{\sqrt{p_0^2-m^2}}+\text{O}(\Delta^2),
\eea
so that the propagator if Eq.~\eqref{eq:Prop_Delta} is:

\bea
    \ii S_\Delta^{(\pm)}(p)&=&\left[\ii\frac{e^{-\pt^2/\qB}}{\pp^2-m^2}+\ii\Delta\left(\frac{3\qB}{4\pi}\right)\frac{(\pp^2+m^2)e^{-2\pt^2/\qB}}{(\pp^2-m^2)^2\sqrt{p_0^2-m^2}}\right]\nn\\
    &\times&\left[(\slashed{p}_\parallel+m)-\Delta \left(\frac{3\qB}{4\pi}\right)\frac{e^{-\pt^2/\qB}}{\sqrt{p_0^2-m^2}}\gamma_3p^3\right]\left[2\mathcal{O}^{(\pm)}\mp\ii\Delta\left(\frac{\qB}{2\pi}\right)\frac{e^{-\pt^2/\qB}}{\sqrt{p_0^2-m^2}}\gamma^1\gamma^2\right]\nn\\
    &=&\left[\ii\frac{e^{-\pt^2/\qB}}{\pp^2-m^2}+\ii\Delta\left(\frac{3\qB}{4\pi}\right)\frac{(\pp^2+m^2)e^{-2\pt^2/\qB}}{(\pp^2-m^2)^2\sqrt{p_0^2-m^2}}\right]\nn\\
    &\times&\left[2(\slashed{p}_\parallel+m)\mathcal{O}^{(\pm)}\mp\ii\Delta\left(\frac{\qB}{2\pi}\right)\frac{e^{-\pt^2/\qB}}{\sqrt{p_0^2-m^2}}(\slashed{p}_\parallel+m)\gamma^1\gamma^2-\Delta \left(\frac{3\qB}{2\pi}\right)\frac{e^{-\pt^2/\qB}}{\sqrt{p_0^2-m^2}}p_3\gamma^3\mathcal{O}^{(\pm)}+\text{O}(\Delta^2)\right]\nn\\
    &=&\ii S_0^{(\pm)}(p)\mp\ii\Delta\left(\frac{\qB}{2\pi}\right)\frac{e^{-2\pt^2/\qB}}{(\pp^2-m^2)\sqrt{p_0^2-m^2}}(\slashed{p}_\parallel+m)\ii\gamma^1\gamma^2-\ii\Delta \left(\frac{3\qB}{2\pi}\right)\frac{e^{-2\pt^2/\qB}}{(\pp^2-m^2)\sqrt{p_0^2-m^2}}p_3\gamma^3\mathcal{O}^{(\pm)}\nn\\
    &+&\ii\Delta\left(\frac{3\qB}{2\pi}\right)\frac{(\pp^2+m^2)e^{-2\pt^2/\qB}}{(\pp^2-m^2)^2\sqrt{p_0^2-m^2}}(\slashed{p}_\parallel+m)\mathcal{O}^{(\pm)}+\text{O}(\Delta^2),
\eea
where
\bea
\ii S_0^{(\pm)}(p)&=&2\ii\frac{e^{-\pt^2/\qB}}{\pp^2-m^2}(\slashed{p}_\parallel+m)\mathcal{O}^{(\pm)}\nn\\
\mathcal{O}^{(\pm)}&=&\frac{1}{2}\left(1\pm\ii\gamma^1\gamma^2\right).
\eea

Let us define
\begin{subequations}
\bea
\Theta_1(p)&\equiv&\frac{3(\pp^2+m^2)e^{-2\pt^2/\qB}}{(\pp^2-m^2)^2\sqrt{p_0^2-m^2}} ,
\eea
\bea
\Theta_2(p)&\equiv&\frac{3p_3e^{-2\pt^2/\qB}}{(\pp^2-m^2)\sqrt{p_0^2-m^2}},
\eea
and
\bea
\Theta_3(p)&\equiv&\frac{e^{-2\pt^2/\qB}}{(\pp^2-m^2)\sqrt{p_0^2-m^2}},
\eea
\end{subequations}
so that the propagator is
\bea
\ii S_\Delta^{(\pm)}(p)&=&\ii S_0^{(\pm)}(p)+\ii\Delta\left(\frac{\qB}{2\pi}\right)\Big[\Theta_1(p)(\slashed{p}_\parallel+m)\mathcal{O}^{(\pm)}-\Theta_2(p)\gamma^3\mathcal{O}^{(\pm)}\pm\Theta_3(p)\ii\gamma^1\gamma^2(\slashed{p}_\parallel+m)\Big]+\text{O}(\Delta^2).
\eea

Therefore, the polarization tensor at order $\text{O}(\Delta^2)$ reads:
\bea
\ii\Pi_\Delta^{\mu\nu}&=&\ii\Pi_0^{\mu\nu}+\ii\frac{q^2\qB\Delta}{4\pi}\sum_{i=1}^{12}\int\frac{d^4k}{(2\pi)^4}t_i^{\mu\nu}(k),
\eea
where $\ii\Pi_0^{\mu\nu}$ is the one-loop polarization tensor in the strong field limit and in the absence of fluctuations~\cite{PhysRevD.83.111501}, and 
\begin{subequations}
    \bea
    t_1^{\mu\nu}=\Theta_1(k-p)\Tr{\gn\ii S_0^{(-)}(k)\gm (\slashed{k}_\parallel-\slashed{p}_\parallel+m)\Op{-}},
    \label{t1}
    \eea
    \bea
    t_2^{\mu\nu}=-\Theta_2(k-p)\Tr{\gn\ii S_0^{(-)}(k)\gm \gamma^3\mathcal{O}^{(-)}},
    \label{t2}
    \eea
    \bea
    t_3^{\mu\nu}=-\ii\Theta_3(k-p)\Tr{\gn\ii S_0^{(-)}(k)\gm(\slashed{k}_\parallel-\slashed{p}_\parallel+m)\gamma^1\gamma^2},
    \label{t3}
    \eea
    \bea
    t_4^{\mu\nu}=\Theta_1(k)\Tr{\gn(\slashed{k}_\parallel+m)\Op{-}\gm \ii S_0^{(-)}(k-p)},
    \label{t4}
    \eea
    \bea
    t_5^{\mu\nu}=-\Theta_2(k)\Tr{\gn\gamma^3\Op{-}\gm \ii S_0^{(-)}(k-p)},
    \label{t5}
    \eea
    \bea
    t_6^{\mu\nu}=-\ii\Theta_3(k)\Tr{\gn(\slashed{k}_\parallel+m)\gamma^1\gamma^2\gm \ii S_0^{(-)}(k-p)},
    \label{t6}
    \eea
    \bea
    t_7^{\mu\nu}=\Theta_1(-k)\Tr{\gn\ii S_0^{(+)}(-k+p)\gm(-\slashed{k}+m)\Op{+}},
    \label{t7}
    \eea
    \bea
    t_8^{\mu\nu}=-\Theta_2(-k)\Tr{\gn\ii S_0^{(+)}(-k+p)\gm\gamma^3\Op{+}},
    \label{t8}
    \eea
    \bea
    t_9^{\mu\nu}=\ii\Theta_3(-k)\Tr{\gn\ii S_0^{(+)}(-k+p)\gm(-\slashed{k}_\parallel+m)\gamma^1\gamma^2},
    \label{t9}
    \eea
    \bea
    t_{10}^{\mu\nu}=\Theta_1(-k+p)\Tr{\gn(-\slashed{k}_\parallel+\slashed{p}_\parallel+m)\Op{+}\gm\ii S_0^{(+)}(-k)},
    \label{t10}
    \eea
    \bea
    t_{11}^{\mu\nu}=-\Theta_2(-k+p)\Tr{\gn\gamma^3\Op{+}\gm\ii S_0^{(+)}(-k)},
    \label{t11}
    \eea
    \bea
    t_{12}^{\mu\nu}=\ii\Theta_3(-k+p)\Tr{\gn\gamma^1\gamma^2\gm(-\slashed{k}_\parallel+\slashed{p}_\parallel+m)\ii S_0^{(+)}(-k)}.
    \label{t12}
    \eea
\end{subequations}

We can add the similar terms:
\bea
&&\T{1}+\T{4}+\T{7}+\T{10}\nn\\
&=&\Theta_1(k-p)\Tr{\gn\ii S_0^{(-)}(k)\gm (\slashed{k}_\parallel-\slashed{p}_\parallel+m)\Op{-}}+\Theta_1(k)\Tr{\gn(\slashed{k}_\parallel+m)\Op{-}\gm \ii S_0^{(-)}(k-p)}\nn\\
&+&\Theta_1(-k)\Tr{\gn\ii S_0^{(+)}(-k+p)\gm(-\slashed{k}+m)\Op{+}}+\Theta_1(-k+p)\Tr{\gn(-\slashed{k}_\parallel+\slashed{p}_\parallel+m)\Op{+}\gm\ii S_0^{(+)}(-k)},
\eea
given that $\Theta_1(-p)=\Theta_1(p)$, we get
\bea
&&\T{1}+\T{4}+\T{7}+\T{10}\nn\\
&=&\Bigg[\frac{2\ii e^{-\kt^2/\qB}}{\kp^2-m^2}\Theta_1(k-p)+\frac{2\ii e^{-(k-p)^2_\perp/\qB}}{(k-p)_\parallel^2-m^2}\Theta_1(k)\Bigg]\nn\\
&\times&\Big[\Tr{\gn(\slashed{k}_\parallel+m)\Op{-}\gm (\slashed{k}_\parallel-\slashed{p}_\parallel+m)\Op{-}}+\Tr{\gn(-\slashed{k}_\parallel+\slashed{p}_\parallel+m)\Op{+}\gm(-\slashed{k}_\parallel+m)\Op{+}}\Big].
\eea

Now, from the identities $\left[\gm_\parallel,\Op{\pm}\right]=0$, and $\Op{\pm}\gm\Op{\pm}=\Op{\pm}\gm_\parallel$, it is straightforward to show that:
\bea
&&\T{1}+\T{4}+\T{7}+\T{10}\nn\\
&=&8\ii\Bigg[\frac{e^{-\kt^2/\qB}}{\kp^2-m^2}\Theta_1(k-p)+\frac{e^{-(k-p)^2_\perp/\qB}}{(k-p)_\parallel^2-m^2}\Theta_1(k)\Bigg]\left[\left(\pp\cdot\kp-\kp^2+m^2\right)g_\parallel^{\mu\nu}+2\kp^\mu\kp^\nu-\pp^\mu\kp^\nu-\pp^\nu\kp^\mu\right]\nn\\
\label{T1}
\eea

Similarly, we can add the following tensors:
\bea
&&\T{2}+\T{5}+\T{8}+\T{11}\nn\\
&=&-\Theta_2(k-p)\Tr{\gn\ii S_0^{(-)}(k)\gm \gamma^3\mathcal{O}^{(-)}}-\Theta_2(k)\Tr{\gn\gamma^3\Op{-}\gm \ii S_0^{(-)}(k-p)}\nn\\
&-&\Theta_2(-k)\Tr{\gn\ii S_0^{(+)}(-k+p)\gm\gamma^3\Op{+}}-\Theta_2(-k+p)\Tr{\gn\gamma^3\Op{+}\gm\ii S_0^{(+)}(-k)}.
\eea

In this case, from the fact that $\Theta_2(-p)=-\Theta_2(p)$:
\bea
&&\T{2}+\T{5}+\T{8}+\T{11}\nn\\
&=&\frac{2\ii e^{-\kt^2/\qB}}{\kp^2-m^2}\Theta_2(k-p)\Bigg[\Tr{\gn\gamma^3\Op{+}\gm(-\slashed{k}_\parallel+m)\Op{+}}-\Tr{\gn(\slashed{k}_\parallel+m)\Op{-}\gm \gamma^3\mathcal{O}^{(-)}}\Bigg]\nn\\
&+&\frac{2\ii e^{-(k-p)^2_\perp/\qB}}{(k-p)_\parallel^2-m^2}\Theta_2(k)\Bigg[\Tr{\gn(-\slashed{k}_\parallel+\slashed{p}_\parallel+m)\Op{+}\gm\gamma^3\Op{+}}-\Tr{\gn\gamma^3\Op{-}\gm (\slashed{k}_\parallel-\slashed{p}_\parallel+m)\Op{-}}\Bigg]\nn\\
&=&8\ii\left[\frac{ e^{-\kt^2/\qB}}{\kp^2-m^2}\Theta_2(k-p)+\frac{e^{-(k-p)^2_\perp/\qB}}{(k-p)_\parallel^2-m^2}\Theta_2(k)\right]\left(k^3g_\parallel^{\mu\nu}+\kp^\mu\delta_3^\nu+\kp^\nu\delta_3^\mu\right)\nn\\
&-&8\ii\frac{e^{-(k-p)^2_\perp/\qB}}{(k-p)_\parallel^2-m^2}\Theta_2(k)\left(p^3g_\parallel^{\mu\nu}+\pp^\mu\delta_3^\nu+\pp^\nu\delta_3^\mu\right).
\label{T2}
\eea

The last group of tensors is:
\bea
&&\T{3}+\T{6}+\T{9}+\T{12}\nn\\
&=&-\ii\Theta_3(k-p)\Tr{\gn\ii S_0^{(-)}(k)\gm (\slashed{k}_\parallel-\slashed{p}_\parallel+m)\gamma^1\gamma^2}-\ii\Theta_3(k)\Tr{\gn(\slashed{k}_\parallel+m)\gamma^1\gamma^2\gm \ii S_0^{(-)}(k-p)}\nn\\
&+&\ii\Theta_3(-k)\Tr{\gn\ii S_0^{(+)}(-k+p)\gm(-\slashed{k}_\parallel+m)\gamma^1\gamma^2}+\ii\Theta_3(-k+p)\Tr{\gn(-\slashed{k}_\parallel+\slashed{p}_\parallel+m)\gamma^1\gamma^2\gm\ii S_0^{(+)}(-k)}.\nn\\
\eea

Because $\Theta_3(-p)=\Theta_3(p)$, we get:
{\small
\bea
&&\T{3}+\T{6}+\T{9}+\T{12}\nn\\
&=&\frac{2 e^{-\kt^2/\qB}}{\kp^2-m^2}\Theta_3(k-p)\Bigg[\Tr{\gn(\slashed{k}_\parallel+m)\Op{-}\gm(\slashed{k}_\parallel-\slashed{p}_\parallel+m) \gamma^1\gamma^2}-\Tr{\gn(-\slashed{k}_\parallel+\slashed{p}_\parallel+m)\gamma^1\gamma^2\gm(-\slashed{k}_\parallel+m)\Op{+}}\Bigg]\nn\\
&+&\frac{2 e^{-(k-p)^2_\perp/\qB}}{(k-p)_\parallel^2-m^2}\Theta_3(k)\Bigg[\Tr{\gn(\slashed{k}_\parallel+m)\gamma^1\gamma^2\gm (\slashed{k}_\parallel-\slashed{p}_\parallel+m)\Op{-}}-\Tr{\gn(-\slashed{k}_\parallel+\slashed{p}_\parallel+m)\Op{+}\gm(-\slashed{k}_\parallel+m)\gamma^1\gamma^2}\Bigg]\nn\\
&=&\frac{e^{-\kt^2/\qB}}{\kp^2-m^2}\Theta_3(k-p)\Bigg[8\epsilon_{ab}g_\perp^{a\mu}g_\perp^{b\nu}\left(m^2+\pp\cdot\kp-\kp^2\right)-\ii\Tr{\gn\slashed{k}_\parallel\gamma^1\gamma^2\gm\gamma^1\gamma^2(\slashed{k}_\parallel-\slashed{p}_\parallel)}\nn\\
&+&\ii\Tr{\gn(-\slashed{k}_\parallel+\slashed{p}_\parallel)\gamma^1\gamma^2\gm\slashed{k}_\parallel\gamma^1\gamma^2}-2\ii m^2\Tr{\gn\gamma^1\gamma^2\gm\gamma^1\gamma^2}\Bigg]\nn\\
&+&\frac{e^{-(k-p)^2_\perp/\qB}}{(k-p)_\parallel^2-m^2}\Theta_3(k)\Bigg[-8\epsilon_{ab}g_\perp^{a\mu}g_\perp^{b\nu}\left(m^2+\pp\cdot\kp-\kp^2\right)-\ii\Tr{\gn\slashed{k}_\parallel\gamma^1\gamma^2\gm(\slashed{k}_\parallel-\slashed{p}_\parallel)\gamma^1\gamma^2}\nn\\
&+&\ii\Tr{\gn(-\slashed{k}_\parallel+\slashed{p}_\parallel)\gamma^1\gamma^2\gm\slashed{k}_\parallel\gamma^1\gamma^2}-2\ii m^2\Tr{\gn\gamma^1\gamma^2\gm\gamma^1\gamma^2}\Bigg]
\eea
}

From the identity: $\gamma^1\gamma^2\gm\gamma^1\gamma^2=\gm_\perp-\gm_\parallel$, we obtain:
\bea
&&\T{3}+\T{6}+\T{9}+\T{12}\nn\\
&=&\frac{8\ii e^{-\kt^2/\qB}}{\kp^2-m^2}\Theta_3(k-p)\Bigg[(m^2+\pp\cdot\kp-\kp^2)(g_\parallel^{\mu\nu}-g_\perp^{\mu\nu}-\ii\epsilon_{ab}g_\perp^{a\mu}g_\perp^{b\nu})+2\kp^\mu\kp^\nu-\pp^\mu\kp^\nu-\pp^\nu\kp^\mu\Bigg]\nn\\
&+&\frac{8\ii e^{-(k-p)^2_\perp/\qB}}{(k-p)_\parallel^2-m^2}\Theta_3(k)\Bigg[(m^2+\pp\cdot\kp-\kp^2)(g_\parallel^{\mu\nu}-g_\perp^{\mu\nu}+\ii\epsilon_{ab}g_\perp^{a\mu}g_\perp^{b\nu})+2\kp^\mu\kp^\nu-\pp^\mu\kp^\nu-\pp^\nu\kp^\mu\Bigg].
\eea

\subsection{The integral $\mathcal{I}(p_{\parallel}^2)$}
Here we present the details on the calculation of the integral
\bea
\mathcal{I}(\pp^2)&=&\int_0^1dx\frac{x(1-x)}{x(1-x)\pp^2-m^2}\nn\\
&=& \frac{1}{\pp^2}\left[ 
1 + \frac{m^2}{\pp^2}\int_0^1 \frac{dx}{x(1 - x) - m^2/\pp^2}
\right].
\eea

The denominator of the remaining integral is a second-order polynomial in $x$, whose roots (poles) are
\be
x = a_{\pm} = \frac{1}{2} \pm \frac{1}{2}\sqrt{1 - \frac{4 m^2}{\pp^2}}.
\ee

Factorizing the denominator accordingly, we have
\bea
\mathcal{I}(\pp^2) &=& \frac{1}{\pp^2}\left[ 
1 - \frac{m^2}{\pp^2}\int_0^1 \frac{dx}{(x - a_{+})(x - a_{-})}
\right]\nn\\
&=& \frac{1}{\pp^2}\left[ 
1 - \frac{m^2}{\pp^2(a_{+} - a_{-})}\int_0^1 dx\left(\frac{1}{x - a_{+}} - \frac{1}{x - a_{-}}\right)
\right]\nn\\
&=& \frac{1}{\pp^2}\left[ 
1 - \frac{m^2}{\pp^2(a_{+} - a_{-})}\log\left[\frac{(1 - a_+)}{(1-a_-)}\frac{a_-}{a_+} \right]
\right].
\eea

Finally, substituting the definitions of $a_{\pm}$, we obtain
\bea
\mathcal{I}(\pp^2)&=&\int_0^1dx\frac{x(1-x)}{x(1-x)\pp^2-m^2}\\
&=& \frac{1}{p_{\parallel}^2}\left[ 1 + \frac{2 m^2/p_{\parallel}^2}{\sqrt{1 - \frac{4 m^2}{p_{\parallel}^2}}}\log\left[ \frac{1 + \sqrt{1 - \frac{4 m^2}{p_{\parallel}^2}}}{1 - \sqrt{1 - \frac{4 m^2}{p_{\parallel}^2}}} \right]  \right].\nn
\eea

\section{Momentum integrals}\label{sec:momentumintegrals}

Let us define the following tensors:
\begin{subequations}
  \bea
T_1^{\mu\nu}\equiv\int\frac{d^4k}{(2\pi)^4}\left(\T{1}+\T{4}+\T{7}+\T{10}\right),
\eea
\bea
T_2^{\mu\nu}&\equiv&\int\frac{d^4k}{(2\pi)^4}\left(\T{2}+\T{5}+\T{8}+\T{11}\right),
\eea
and
\bea
T_3^{\mu\nu}&\equiv&\int\frac{d^4k}{(2\pi)^4}\left(\T{3}+\T{6}+\T{9}+\T{12}\right)
\eea
\end{subequations}

\subsection{Computing $T_1^{\mu\nu}$}
\bea
&&T_1^{\mu\nu}\equiv\int\frac{d^4k}{(2\pi)^4}\left(\T{1}+\T{4}+\T{7}+\T{10}\right)\nn\\
&=&8\ii\int\frac{d^4k}{(2\pi)^4}\Bigg[\frac{e^{-\kt^2/\qB}}{\kp^2-m^2}\Theta_1(k-p)+\frac{e^{-(k-p)^2_\perp/\qB}}{(k-p)_\parallel^2-m^2}\Theta_1(k)\Bigg]\left[\left(\pp\cdot\kp-\kp^2+m^2\right)g_\parallel^{\mu\nu}+2\kp^\mu\kp^\nu-\pp^\mu\kp^\nu-\pp^\nu\kp^\mu\right].\nn\\
&=&T_{1(\text{a})}^{\mu\nu}+T_{1(\text{b})}^{\mu\nu},
\eea
where
\begin{subequations}
  \bea
T_{1(\text{a})}^{\mu\nu}&\equiv&8\ii\int\frac{d^4k}{(2\pi)^4}\frac{e^{-\kt^2/\qB}}{\kp^2-m^2}\Theta_1(k-p)\left[\left(\kp\cdot(\pp-\kp)+m^2\right)g_\parallel^{\mu\nu}+(\kp^\mu - \pp^\mu)\kp^\nu+\kp^\mu(\kp^\nu-\pp^\nu)\right]
\eea
and
\bea
T_{1(\text{b})}^{\mu\nu}&\equiv&8\ii\int\frac{d^4k}{(2\pi)^4}\frac{e^{-(k-p)^2_\perp/\qB}}{(k-p)_\parallel^2-m^2}\Theta_1(k)\left[\left(\kp\cdot(\pp-\kp)+m^2\right)g_\parallel^{\mu\nu}+(\kp^\mu - \pp^\mu)\kp^\nu+\kp^\mu(\kp^\nu-\pp^\nu)\right].
\eea
\end{subequations}
In the second expression for $T_{1(\text{b})}^{\mu\nu}$, let us change the integration variable $k' = p - k$, such that $k = p - k'$ and $d^{4}k' = d^{4}k$,
\bea
T_{1(\text{b})}^{\mu\nu}&\equiv&8\ii\int\frac{d^4k'}{(2\pi)^4}\frac{e^{-k'^2_\perp/\qB}}{\kp'^2-m^2}\Theta_1(p-k')\left[\left((\pp - \kp')\cdot\kp'+m^2\right)g_\parallel^{\mu\nu}+(-\kp'^\mu )(\pp^\nu - \kp'^\nu)+(\pp^\mu-\kp'^\mu)(-\kp'^\nu)\right].
\eea
Finally, using the parity symmetry of the function $\Theta_1(p - k') = \Theta_1(k'-p)$, and removing the primes, we obtain
\bea
T_{1(\text{b})}^{\mu\nu}&\equiv&8\ii\int\frac{d^4k}{(2\pi)^4}\frac{e^{-k^2_\perp/\qB}}{\kp^2-m^2}\Theta_1(k-p)\left[\left((\pp - \kp)\cdot\kp+m^2\right)g_\parallel^{\mu\nu}+\kp^\mu(\kp^\nu - \pp^\nu)+(\kp^\mu-\pp^\mu)\kp^\nu\right],
\eea
to conclude that $T_{1(\text{b})}^{\mu\nu} = T_{1(\text{a})}^{\mu\nu}$, and hence $T_{1}^{\mu\nu} = 2T_{1(\text{a})}^{\mu\nu}$.

For both tensors the integration over the perpendicular momenta is the same. For $T_{1(\text{a})}^{\mu\nu}$:
\bea
\frac{e^{-\kt^2/\qB}}{\kp^2-m^2}\Theta_1(k-p)=\frac{3\left[(k-p)_\parallel^2+m^2\right]e^{-\kt^2/\qB}e^{2(\kt-\pt)^2/\qB}}{(\kp^2-m^2)\left[(k-p)_\parallel^2-m^2\right]^2\sqrt{(k_0-p_0)^2-m^2}},
\eea
where the factors in the exponential can be reduced as follows:
\bea
\kt^2+2(\kt-\pt)^2=3\kt^2-4\pt\cdot\kt+2\pt^2=3\left(\kt-2\pt/3\right)^2+2\pt^2/3.
\eea

Then, the suggested change of variables is given by 
\bea
\ell_\perp^\mu\equiv \kt^\mu-\frac{2}{3}\pt^\mu,
\eea
so that the integration over perpendicular momenta is straightforward:
\begin{subequations}
    \bea
T_{1(\text{a})}^{\mu\nu}&=&\frac{8\ii\pi\qB}{(2\pi)^2} \exp\left(-\frac{2\pt^2}{3\qB}\right)\int\frac{d^2k_\parallel}{(2\pi)^2}\frac{\left[(k-p)_\parallel^2+m^2\right]\left[\left(\pp\cdot\kp-\kp^2+m^2\right)g_\parallel^{\mu\nu}+2\kp^\mu\kp^\nu-\pp^\mu\kp^\nu-\pp^\nu\kp^\mu\right]}{(\kp^2-m^2)\left[(k-p)_\parallel^2-m^2\right]^2\sqrt{(k_0-p_0)^2-m^2}}.
\eea
%
\end{subequations}

From the identity:
\bea
\frac{1}{AB^2}=2\int_0^\infty\frac{\lambda~d\lambda}{(\lambda A+B)^3},
\eea
denominator of $T_{1(\text{a})}^{\mu\nu}$ is
\bea
\frac{1}{(\kp^2-m^2)\left[(k-p)_\parallel^2-m^2\right]^2}&=&2\int_0^\infty\frac{\lambda~d\lambda}{\left[(1+\lambda)\kp^2-2\pp\cdot\kp+\pp^2-(1+\lambda)m^2\right]^3}\nn\\
&=&2\int_0^\infty\frac{\lambda~d\lambda}{\left[(1+\lambda)\left(\kp-\frac{1}{1+\lambda}\pp\right)^2+\frac{\lambda}{1+\lambda}\pp^2-(1+\lambda)m^2\right]^3},
\eea
which suggest that the shift in the parallel momenta must be
\bea
\lp^\mu=\kp^\mu-\frac{1}{1+\lambda}\pp^\mu. 
\eea

Similarly:
\bea
\frac{1}{[(k-p)_\parallel^2-m^2](k_\parallel^2-m^2)^2}=2\int_0^\infty\frac{\lambda~d\lambda}{\left[(1+\lambda)\left(\kp-\frac{\lambda}{1+\lambda}\pp\right)^2+\frac{\lambda}{1+\lambda}\pp^2-(1+\lambda)m^2\right]^3},
\eea
so that
\bea
\lp^\mu=\kp^\mu-\frac{\lambda}{1+\lambda}\pp^\mu. 
\eea

Then
\begin{subequations}
  \bea
T_{1(\text{a})}^{\mu\nu}&=&\frac{16\ii\pi\qB}{(2\pi)^2} \exp\left(-\frac{2\pt^2}{3\qB}\right)\int_0^\infty\frac{\lambda d\lambda}{(1+\lambda)^3}\int\frac{d^2\ell_\parallel}{(2\pi)^2}\frac{\left(\lp-\frac{\lambda}{1+\lambda}\pp\right)^2+m^2}{\left(\ell_\parallel^2+\frac{\lambda}{(1+\lambda)^2}\pp^2-m^2\right)^3\sqrt{\left(l_0-\frac{\lambda}{1+\lambda}p_0\right)^2-m^2}}\nn\\
&\times&\left[\left(\frac{\lambda-1}{1+\lambda}\pp\cdot\lp+\frac{\lambda}{(1+\lambda)^2}\pp^2-\lp^2+m^2\right)g_\parallel^{\mu\nu}+2\lp^\mu\lp^\nu+\frac{1-\lambda}{1+\lambda}(\pp^\mu\lp^\nu+\pp^\nu\lp^\mu)-\frac{2\lambda}{(1+\lambda)^2}\pp^\mu\pp^\nu\right].
\eea
%
\end{subequations}
%
%

By ignoring odd powers on $\lp$, and by using the fact that under the integral
\bea
\lp^\mu\lp^\nu\to\frac{1}{2}\lp^2g_\parallel^{\mu\nu}
\eea
the expressions reduce to
\bea
T_{1(\text{a})}^{\mu\nu}&=&\frac{16\ii\pi\qB}{(2\pi)^2} \exp\left(-\frac{2\pt^2}{3\qB}\right)\int_0^\infty\frac{\lambda d\lambda}{(1+\lambda)^3}\int\frac{d^2\ell_\parallel}{(2\pi)^2}\frac{1}{\left(\ell_\parallel^2+\frac{\lambda}{(1+\lambda)^2}\pp^2-m^2\right)^3\sqrt{\left(l_0-\frac{\lambda}{1+\lambda}p_0\right)^2-m^2}}\nn\\
&\times&\left\{\left(\lp^2+\frac{\lambda^2}{(1+\lambda)^2}\pp^2+m^2\right)\left[\left(\frac{\lambda}{(1+\lambda)^2}\pp^2+m^2\right)\gp^{\mu\nu}-\frac{2\lambda}{(1+\lambda)^2}\pp^\mu\pp^\nu\right]\right.\nn\\
&+&\left.\frac{2\lambda(1-\lambda)}{(1+\lambda)^2}(\pp\cdot\lp)\left[(\pp\cdot\lp)\gp^{\mu\nu}-(\pp^\mu\lp^\nu+\pp^\nu\lp^\mu)\right]\right\}.
\eea

Now, under the integral we can replace:
\bea
(\pp\cdot\lp)(\pp\cdot\lp)&\to&\frac{1}{2}\pp^2\lp^2\nn\\
(\pp\cdot\lp)\pp^\mu\lp^\nu&\to&\frac{1}{2}\lp^2\pp^\mu\pp^\nu,
\eea
therefore,
\bea
T_{1(\text{a})}^{\mu\nu}&=&\frac{16\ii\pi\qB}{(2\pi)^2} \exp\left(-\frac{2\pt^2}{3\qB}\right)\int_0^\infty\frac{\lambda d\lambda}{(1+\lambda)^3}\int\frac{d^2\ell_\parallel}{(2\pi)^2}\frac{1}{\left(\ell_\parallel^2+\frac{\lambda}{(1+\lambda)^2}\pp^2-m^2\right)^3\sqrt{\left(l_0-\frac{\lambda}{1+\lambda}p_0\right)^2-m^2}}\nn\\
&\times&\left\{\left(\lp^2+\frac{\lambda^2}{(1+\lambda)^2}\pp^2+m^2\right)\left[\frac{\lambda}{(1+\lambda)^2}\left(\pp^2\gp^{\mu\nu}-2\pp^\mu\pp^\nu\right)+m^2\gp^{\mu\nu}\right]+\frac{\lambda(1-\lambda)}{(1+\lambda)^2}\lp^2\left(\pp^2\gp^{\mu\nu}-2\pp^\mu\pp^\nu\right)\right\},
\eea
and $T_{1}^{\mu\nu} = 2 T_{1(a)}^{\mu\nu}$.
\subsection{Computing $T_2^{\mu\nu}$}
%
\bea
T_2^{\mu\nu}&\equiv&\int\frac{d^4k}{(2\pi)^4}\left(\T{2}+\T{5}+\T{8}+\T{11}\right)\nn\\
&=&8\ii\int\frac{d^4k}{(2\pi)^4}\left[\frac{ e^{-\kt^2/\qB}}{\kp^2-m^2}\Theta_2(k-p)+\frac{e^{-(k-p)^2_\perp/\qB}}{(k-p)_\parallel^2-m^2}\Theta_2(k)\right]\left(k^3g_\parallel^{\mu\nu}+\kp^\mu\delta_3^\nu+\kp^\nu\delta_3^\mu\right)\nn\\
&-&8\ii\int\frac{d^4k}{(2\pi)^4}\frac{e^{-(k-p)^2_\perp/\qB}}{(k-p)_\parallel^2-m^2}\Theta_2(k)\left(p^3g_\parallel^{\mu\nu}+\pp^\mu\delta_3^\nu+\pp^\nu\delta_3^\mu\right)\nn\\
&=&T_{2(\text{a})}^{\mu\nu}+T_{2(\text{b})}^{\mu\nu}
\eea
with
\begin{subequations}
    \bea
    T_{2(\text{a})}^{\mu\nu}&\equiv&8\ii\int\frac{d^4k}{(2\pi)^4}\frac{ e^{-\kt^2/\qB}}{\kp^2-m^2}\Theta_2(k-p)\left(k^3g_\parallel^{\mu\nu}+\kp^\mu\delta_3^\nu+\kp^\nu\delta_3^\mu\right)\nn\\
    &=&\frac{8\ii\pi\qB}{(2\pi)^2}\exp\left(-\frac{2\pt^2}{3\qB}\right)\int\frac{d^2\kp}{(2\pi)^2}\frac{(k_3-p_3)\left(k^3g_\parallel^{\mu\nu}+\kp^\mu\delta_3^\nu+\kp^\nu\delta_3^\mu\right)}{(\kp^2-m^2)[(k-p)_\parallel^2-m^2]\sqrt{(k_0-p_0)^2-m^2}}
    \eea
    \bea
    T_{2(\text{b})}^{\mu\nu}&\equiv&8\ii\int\frac{d^4k}{(2\pi)^4}\frac{ e^{-(k-p)_\perp^2/\qB}}{(k-p)_\parallel^2-m^2}\Theta_2(k)\left((k^3 - p^3)g_\parallel^{\mu\nu}+(\kp^\mu -\pp^\mu)\delta_3^\nu+(\kp^\nu - \pp^\nu)\delta_3^\mu\right).
    \eea
    %
\end{subequations}

In the expression for $T_{2(\text{b})}^{\mu\nu}$, let us change the integration variable $k' = p - k$, such that $k = p - k'$ and $d^{4}k' = d^{4}k$,
\bea
T_{2(\text{b})}^{\mu\nu}&\equiv&8\ii\int\frac{d^4k}{(2\pi)^4}\frac{ e^{-k'^2_{\perp}/\qB}}{\kp'^2-m^2}\Theta_2(p-k')\left(-k'^3g_\parallel^{\mu\nu}-\kp'^\mu \delta_3^\nu-\kp'^\nu \delta_3^\mu\right)
\eea

Finally, using the odd property of $\Theta_2(p - k') = - \Theta_2(k'-p)$, and removing the primes $k'\rightarrow k$, we finally obtain
\bea
T_{2(\text{b})}^{\mu\nu}&\equiv&8\ii\int\frac{d^4k}{(2\pi)^4}\frac{ e^{-k_\perp^2/\qB}}{k_\parallel^2-m^2}\Theta_2(k-p)\left(k^3g_\parallel^{\mu\nu}+\kp^\mu \delta_3^\nu+\kp^\nu \delta_3^\mu\right)\\
&=& T_{2(\text{a})}^{\mu\nu},
\eea
and hence we conclude $T_{2}^{\mu\nu}=2T_{2(\text{a})}^{\mu\nu}$.

Introducing a Feynman parameter $\lambda$ via the integral transformation
\bea
\frac{1}{AB}=\int_0^\infty\frac{d\lambda}{(\lambda A+B)^2},
\eea
we have for $T_{2(\text{a})}^{\mu\nu}$
\bea
\frac{1}{(\kp^2-m^2)[(k-p)_\parallel^2-m^2]}&=&\int_0^\infty\frac{d\lambda}{\left[\lambda(\kp^2-m^2)+(k-p)_\parallel^2-m^2\right]^2}\nn\\
&=&\int_0^\infty\frac{d\lambda}{(1+\lambda)^2\left[\lp^2+\frac{\lambda}{(1+\lambda)^2}\pp^2-m^2\right]^2}.
\eea
where
\bea
\lp^\mu=\kp^\mu-\frac{1}{1+\lambda}\pp^\mu,
\label{shift1}
\eea
and 
%
\bea
\frac{1}{(\kp^2-m^2)[(k-p)_\parallel^2-m^2]}&=&\int_0^\infty\frac{d\lambda}{\left[\lambda(k-p)_\parallel^2-\lambda m^2+(\kp^2-m^2)\right]^2}\nn\\
&=&\int_0^\infty\frac{d\lambda}{(1+\lambda)^2\left[\lp^2+\frac{\lambda}{(1+\lambda)^2}\pp^2-m^2\right]^2}.
\eea
where
\bea
\lp^\mu=\kp^\mu-\frac{\lambda}{1+\lambda}\pp^\mu,
\label{shift2}
\eea

Therefore, by ignoring odd powers on $\lp$ and using that $p_3p^3=-p_3^2$
\begin{subequations}
    \bea
    T_{2(\text{a})}^{\mu\nu}&=&\frac{8\ii\pi\qB}{(2\pi)^2}\exp\left(-\frac{2\pt^2}{3\qB}\right)\int_0^\infty\frac{d\lambda}{(1+\lambda)^2}\int\frac{d^2\lp}{(2\pi)^2}\frac{\left(-(\ell^3)^2+\frac{\lambda}{(1+\lambda)^2}p_3^2\right)\left(\gp^{\mu\nu}+2\delta_3^\mu\delta_3^\nu\right)}{\left(\lp^2+\frac{\lambda}{(1+\lambda)^2}\pp^2-m^2\right)^2\sqrt{\left(l_0-\frac{\lambda}{1+\lambda}p_0\right)^2-m^2}},
    \eea
    %
    %
\end{subequations}

\subsection{Computing $T_3^{\mu\nu}$}
\bea
&&T_3^{\mu\nu}\equiv\int\frac{d^4k}{(2\pi)^4}\left(\T{3}+\T{6}+\T{9}+\T{12}\right)\nn\\
&=&8\ii \int\frac{d^4k}{(2\pi)^4}\frac{e^{-\kt^2/\qB}}{\kp^2-m^2}\Theta_3(k-p)\Bigg[(m^2+\pp\cdot\kp-\kp^2)(g_\parallel^{\mu\nu}-g_\perp^{\mu\nu}-\ii\epsilon_{ab}g_\perp^{a\mu}g_\perp^{b\nu})+2\kp^\mu\kp^\nu-\pp^\mu\kp^\nu-\pp^\nu\kp^\mu\Bigg]\nn\\
&+&8\ii \int\frac{d^4k}{(2\pi)^4}\frac{e^{-(k-p)^2_\perp/\qB}}{(k-p)_\parallel^2-m^2}\Theta_3(k)\Bigg[(m^2+\pp\cdot\kp-\kp^2)(g_\parallel^{\mu\nu}-g_\perp^{\mu\nu}+\ii\epsilon_{ab}g_\perp^{a\mu}g_\perp^{b\nu})+2\kp^\mu\kp^\nu-\pp^\mu\kp^\nu-\pp^\nu\kp^\mu\Bigg]\nn\\
&=&T_{3(\text{a})}^{\mu\nu}+T_{3(\text{b})}^{\mu\nu}
\eea
where
\begin{subequations}
  \bea
T_{3(\text{a})}^{\mu\nu}&\equiv&8\ii\int\frac{d^4k}{(2\pi)^4}\frac{e^{-\kt^2/\qB}}{\kp^2-m^2}\Theta_3(k-p)\Bigg[(m^2+\kp\cdot(\pp-\kp))(g_\parallel^{\mu\nu}-g_\perp^{\mu\nu}-\ii\epsilon_{ab}g_\perp^{a\mu}g_\perp^{b\nu})+(\kp^\mu - \pp^\mu)\kp^\nu+ \kp^\mu(\kp^\nu - \pp^\nu)\Bigg],\nn\\
\label{eq:T3a}
\eea
and
\bea
T_{3(\text{b})}^{\mu\nu}&\equiv& 8\ii \int\frac{d^4k}{(2\pi)^4}\frac{e^{-(k-p)^2_\perp/\qB}}{(k-p)_\parallel^2-m^2}\Theta_3(k)\Bigg[(m^2+\kp\cdot(\pp-\kp))(g_\parallel^{\mu\nu}-g_\perp^{\mu\nu}+\ii\epsilon_{ab}g_\perp^{a\mu}g_\perp^{b\nu})+(\kp^\mu - \pp^\mu)\kp^\nu+ \kp^\mu(\kp^\nu - \pp^\nu)\Bigg]\nn\\.
\eea
\end{subequations}
We remark that the "skew" terms proportional to $\ii\epsilon_{ab}g_\perp^{a\mu}g_\perp^{b\nu}$ in the expressions above vanish upon integration when they enter in the combination $T_3^{\mu\nu} = T_{3(\text{a})}^{\mu\nu} + T_{3(\text{b})}^{\mu\nu}$. For this purpose,
let us consider the expression for $T_{3(\text{b})}^{\mu\nu}$, and perform the change of integration variables $k' = p - k$, such that $k = p - k'$ and $d^4k' = d^4 k$,
\bea
T_{3(\text{b})}^{\mu\nu}&\equiv&8\ii\int\frac{d^4k'}{(2\pi)^4}\frac{e^{-\kt'^2/\qB}}{\kp'^2-m^2}\Theta_3(p-k')\Bigg[(m^2+(\pp - \kp')\cdot\kp')(g_\parallel^{\mu\nu}-g_\perp^{\mu\nu}+\ii\epsilon_{ab}g_\perp^{a\mu}g_\perp^{b\nu}) -\kp'^\mu(\pp^\nu -\kp'^\nu) + (\pp^\mu -\kp'^\mu)(-\kp'^\nu)\Bigg].\nn\\
\eea
Finally, we make use of the parity property of the function $\Theta_3(p - k') = \Theta_3(k' - p)$, and further removing the primes of the integration variable $k'\rightarrow k$ we obtain 
\bea
T_{3(\text{b})}^{\mu\nu}&\equiv&8\ii\int\frac{d^4k}{(2\pi)^4}\frac{e^{-\kt^2/\qB}}{\kp^2-m^2}\Theta_3(k-p)\Bigg[(m^2+\kp\cdot(\pp - \kp))(g_\parallel^{\mu\nu}-g_\perp^{\mu\nu}+\ii\epsilon_{ab}g_\perp^{a\mu}g_\perp^{b\nu}) +\kp^\mu(\kp^\nu -\pp^\nu) + (\kp^\mu -\pp^\mu)(\kp^\nu)\Bigg].\nn\\
\label{eq:T3btrans}
\eea
Comparing Eq.~\eqref{eq:T3btrans} with Eq.~\eqref{eq:T3a}, we see that they are identical except for the opposite sign of the $\ii\epsilon_{ab}g_\perp^{a\mu}g_\perp^{b\nu}$ terms, that hence vanish upon adding them both, such that
\bea
T_3^{\mu\nu} = 16\ii\int\frac{d^4k}{(2\pi)^4}\frac{e^{-\kt^2/\qB}}{\kp^2-m^2}\Theta_3(k-p)\Bigg[(m^2+\kp\cdot(\pp-\kp))(g_\parallel^{\mu\nu}-g_\perp^{\mu\nu})+(\kp^\mu - \pp^\mu)\kp^\nu+ \kp^\mu(\kp^\nu - \pp^\nu)\Bigg]
\eea
After integrating with respect to perpendicular momenta, we employ the transformation of Eq.~\eqref{shift1} for $T_{3(\text{a})}^{\mu\nu}$ and the transformation of Eq.~\eqref{shift2} for $T_{3(\text{b})}^{\mu\nu}$. Then, ensuring the elimination of odd powers in $\lp$:
\bea
T_{3}^{\mu\nu}&=&\frac{16\ii\pi\qB}{3(2\pi)^2}\exp\left(-\frac{2\pt^2}{3\qB}\right)\int_0^{\infty}\frac{d\lambda}{(1+\lambda)^2}\int\frac{d^2\lp}{(2\pi)^2}\frac{\left(m^2-\lp^2+\frac{\lambda}{(1+\lambda)^2}\pp^2\right)\left(\gp^{\mu\nu}-\gt^{\mu\nu}\right)+\lp^2\gp^{\mu\nu}-\frac{2\lambda}{(1+\lambda)^2}\pp^\mu\pp^\nu}{\sqrt{\left(l_0^2+\frac{\lambda}{1+\lambda}p_0\right)^2-m^2}\left(\lp^2+\frac{\lambda}{(1+\lambda)^2}\pp^2-m^2\right)^2}\nn\\
\eea



%

\section{Integrals in the limit $p_0=0,\mathbf{p}\to0$}\label{sec:Int_pto0}

\subsection{Computing $T_1^{\mu\nu}$}
In the limit $p_0=0,\mathbf{p}\to0$ we have
\bea
T_{1(\text{b})}^{\mu\nu}=T_{1(\text{a})}^{\mu\nu},
\eea
so that
\bea
\lim_{\substack{p_0\to0\\ \mathbf{p}\to0}}T_{1}^{\mu\nu}&=&2T_{1(\text{a})}^{\mu\nu}\nn\\
&=&\frac{32\ii\pi\qB}{(2\pi)^2} \int_0^\infty\frac{\lambda d\lambda}{(1+\lambda)^3}\int\frac{d^2\ell_\parallel}{(2\pi)^2}\frac{m^2(\lp^2+m^2)}{(\ell^2_\parallel-m^2)^3\sqrt{l_0^2-m^2}}\gp^{\mu\nu}\nn\\
&=&\frac{16\ii\pi\qB}{(2\pi)^2} \int\frac{d^2\ell_\parallel}{(2\pi)^2}\frac{m^2(\lp^2+m^2)}{(\ell^2_\parallel-m^2)^3\sqrt{l_0^2-m^2}}\gp^{\mu\nu}.
\eea

Let us define:
\bea
\mathcal{M}_1=\int\frac{d^2\ell_\parallel}{(2\pi)^2}\frac{\ell^2_\parallel+m^2}{\sqrt{\ell_0^2-m^2}(\ell^2_\parallel-m^2)^3}
\eea
so that
\bea
\lim_{\substack{p_0\to0\\ \mathbf{p}\to0}}T_{1}^{\mu\nu}=\frac{16\ii\pi\qB}{(2\pi)^2}m^2\mathcal{M}_1\gp^{\mu\nu}.
\eea

To compute $\mathcal{M}_1$, it is convenient to pass to the Euclidean space $\ell^0\to\ii\ell^4$,
\bea
\mathcal{M}_1=\frac{1}{m^3}\int\frac{dydx}{(2\pi)^2}\frac{y^2+x^2-1}{\sqrt{y^2+1}(y^2+x^2+1)^3}
\eea
where we have defined
\bea
x=\ell^3/m,~\text{and}~y=\ell^4/m.
\label{eq:xandy}
\eea

If $\alpha_\pm^2=y^2\pm1$:
\bea
\mathcal{M}_1=\frac{1}{m^3}\int_{-\infty}^{+\infty}\frac{dy}{2\pi}\frac{1}{\sqrt{y^2+1}}\int_{-\infty}^{+\infty}\frac{dx}{2\pi}\frac{x^2+\alpha_-^2}{(x^2+\alpha_+^2)^3},
\eea
we can perform the $x$-integral of $\mathcal{M}_1$ in the complex plane, so that:
\bea
\int_{-\infty}^{+\infty}\frac{dx}{2\pi}\frac{x^2+\alpha_-^2}{(x^2+\alpha_+^2)^3}=\lim_{R\to\infty}\oint_\mathcal{C}\frac{dz}{2\pi}\frac{z^2+\alpha_-^2}{(z^2+\alpha_+^2)^3}=\ii~\text{Res}\left[\frac{z^2+\alpha_-^2}{(z^2+\alpha_+^2)^3}\right]_{z=\ii\alpha_+}
\eea
where the integration contour $\mathcal{C}$ is shown in Fig.~\ref{fig:M1}(a).

\begin{figure}[h!]
    \centering
    \includegraphics{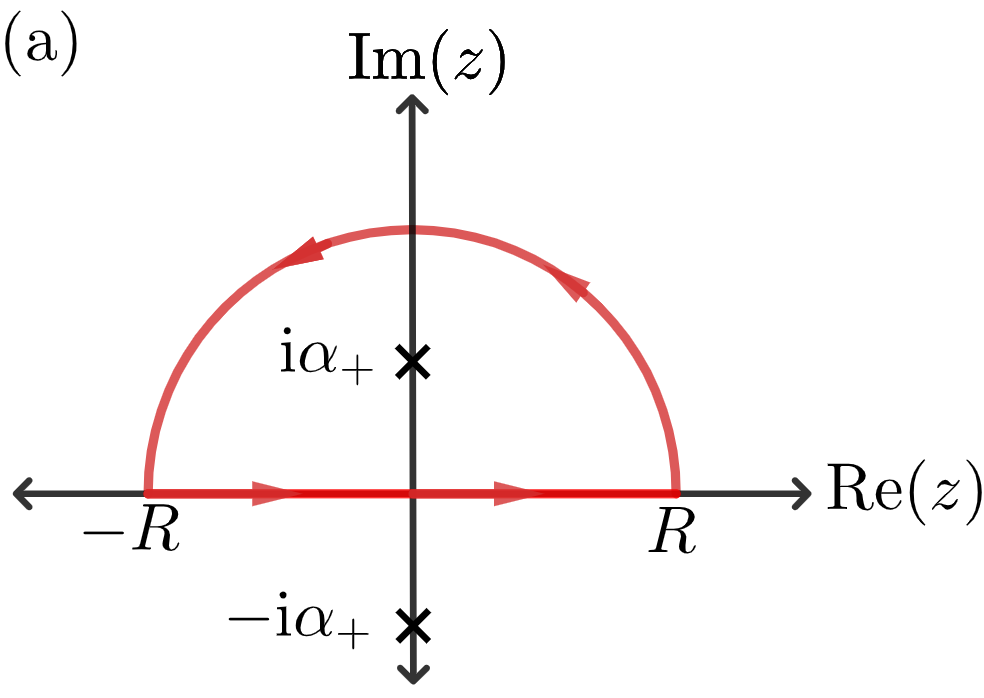}~~~\includegraphics{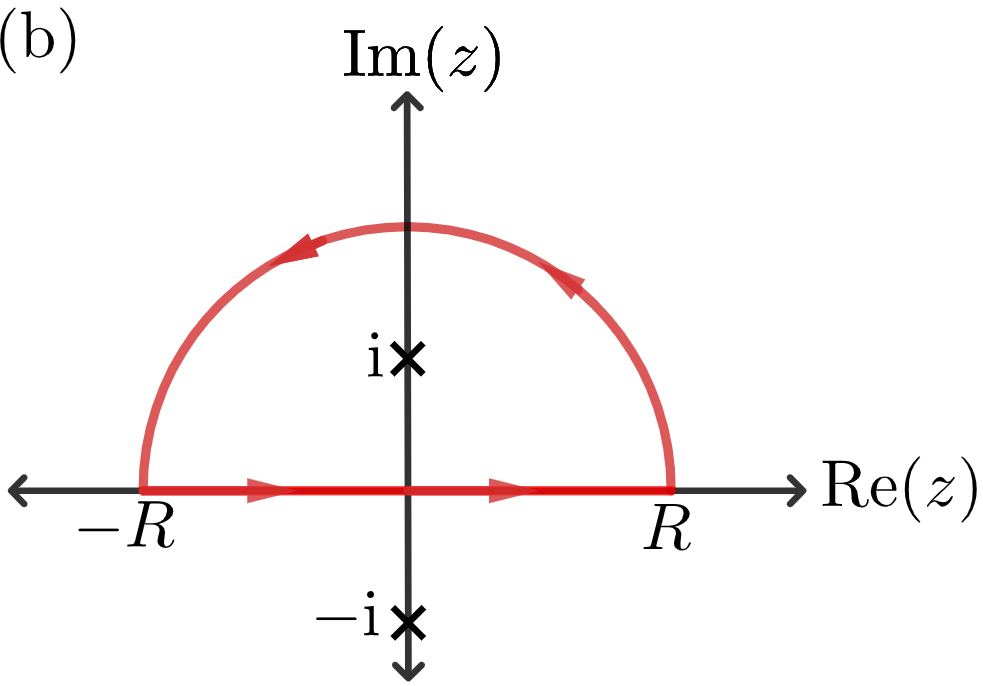}
    \caption{Contours for (a) the $x$-integration of $\mathcal{M}_1$, (b) the $y$-integration of $\mathcal{M}_1$}
    \label{fig:M1}
\end{figure}

Then, it is straightforward to show that
\bea
\mathcal{M}_1=\frac{1}{16m^3}\int_{-\infty}^{+\infty}\frac{dy}{2\pi}\left[\frac{1}{(y^2+1)^2}+\frac{3(y^2-1)}{(y^2+1)^3}\right],
\eea
which can be easily computed with the contour of Fig.~\ref{fig:M1}(b):
\bea
\mathcal{M}_1=-\frac{1}{128m^3}.
\eea

By collecting the results:
\bea
\lim_{\substack{p_0\to0\\ \mathbf{p}\to0}}T_{1}^{\mu\nu}=-\frac{\ii\qB}{32\pi m}\gp^{\mu\nu}.
\eea

\subsection{Computing $T_2^{\mu\nu}$}
In the limit $p_0=0,\mathbf{p}\to0$:
\bea
\lim_{\substack{p_0\to0\\ \mathbf{p}\to0}}T_{2(\text{a})}^{\mu\nu}=\lim_{\substack{p_0\to0\\ \mathbf{p}\to0}}T_{2(\text{b})}^{\mu\nu},~\text{and}~\lim_{\substack{p_0\to0\\ \mathbf{p}\to0}}T_{2(\text{c})}^{\mu\nu}=0,
\eea
then
\bea
\lim_{\substack{p_0\to0\\ \mathbf{p}\to0}}T_{2}^{\mu\nu}&=&-\frac{16\ii\pi\qB}{(2\pi)^2}\int_0^\infty\frac{d\lambda}{(1+\lambda)^2}\int\frac{d^2\lp}{(2\pi)^2}\frac{(\ell^3)^2}{(\lp^2-m^2)^2\sqrt{l_0^2-m^2}}\left(\gp^{\mu\nu}+2\delta_3^\mu\delta_3^\nu\right)\nn\\
&=&-\frac{16\ii\pi\qB}{(2\pi)^2}\int\frac{d^2\lp}{(2\pi)^2}\frac{(\ell^3)^2}{(\lp^2-m^2)^2\sqrt{l_0^2-m^2}}\left(\gp^{\mu\nu}+2\delta_3^\mu\delta_3^\nu\right).
\eea

Passing to the Euclidean space and defining $x$ and $y$ like in Eq.~\eqref{eq:xandy}:
\bea
\lim_{\substack{p_0\to0\\ \mathbf{p}\to0}}T_{2}^{\mu\nu}&=&-\frac{16\ii\pi\qB}{(2\pi)^2m}\int_{-\infty}^{+\infty}\frac{dy}{2\pi}\frac{1}{\sqrt{y^2+1}}\int_{-\infty}^{+\infty}\frac{dx}{2\pi}\frac{x^2}{(x^2+\alpha_+^2)^2}\left(\gp^{\mu\nu}+2\delta_3^\mu\delta_3^\nu\right),
\eea
so that using the integration contour of Fig.~\ref{fig:M1}(a):
\bea
\int_{-\infty}^{+\infty}\frac{dx}{2\pi}\frac{x^2}{(x^2+\alpha_+^2)^2}=\lim_{R\to\infty}\oint_\mathcal{C}\frac{dz}{2\pi}\frac{z^2}{(z^2+\alpha_+^2)^2}=\ii~\text{Res}\left[\frac{z^2}{(z^2+\alpha_+^2)^2}\right]_{z=\ii\alpha_+}=\frac{1}{4\alpha_+}=\frac{1}{4\sqrt{y^2+1}},
\eea
therefore
\bea
\lim_{\substack{p_0\to0\\ \mathbf{p}\to0}}T_{2}^{\mu\nu}&=&-\frac{4\ii\pi\qB}{(2\pi)^2m}\int_{-\infty}^{+\infty}\frac{dy}{2\pi}\frac{1}{(y^2+1)}\left(\gp^{\mu\nu}+2\delta_3^\mu\delta_3^\nu\right),
\eea
and after using the contour of Fig.~\ref{fig:M1}(b):
\bea
\int_{-\infty}^{+\infty}\frac{dy}{2\pi}\frac{1}{(y^2+1)}=\lim_{R\to\infty}\oint_\mathcal{C}\frac{dz}{2\pi}\frac{1}{z^2+1}=\ii~\text{Res}\left[\frac{1}{z^2+1}\right]_{z=\ii}=\frac{1}{2},
\eea
so that
\bea
\lim_{\substack{p_0\to0\\ \mathbf{p}\to0}}T_{2}^{\mu\nu}&=&-\frac{\ii\qB}{2\pi m}\left(\gp^{\mu\nu}+2\delta_3^\mu\delta_3^\nu\right).
\eea

\subsection{Computing $T_3^{\mu\nu}$}
\bea
\lim_{\substack{p_0\to0\\ \mathbf{p}\to0}}T_{3}^{\mu\nu}&=&\frac{16\ii\pi\qB}{3(2\pi)^2}\int_0^{\infty}\frac{d\lambda}{(1+\lambda)^2}\int\frac{d^2\lp}{(2\pi)^2}\frac{\left(m^2-\lp^2\right)\left(\gp^{\mu\nu}-\gt^{\mu\nu}\right)+\lp^2\gp^{\mu\nu}}{(\lp^2-m^2)^2\sqrt{l_0^2-m^2}}\nn\\
&=&\frac{16\ii\pi\qB}{3(2\pi)^2}\int\frac{d^2\lp}{(2\pi)^2}\left[\frac{m^2}{(\lp^2-m^2)^2\sqrt{l_0^2-m^2}}\gp^{\mu\nu}+\frac{\lp^2-m^2}{(\lp^2-m^2)^2\sqrt{l_0^2-m^2}}\gt^{\mu\nu}\right]\nn\\
&=&\frac{16\ii\pi\qB}{3(2\pi)^2}\int\frac{d^2\ell_\text{E}}{(2\pi)^2}\left[\frac{m^2}{(\ell_\text{E}+m^2)^2\sqrt{l_4^2+m^2}}\gp^{\mu\nu}-\frac{\ell_\text{E}^2+m^2}{(\ell_\text{E}+m^2)^2\sqrt{l_4^2+m^2}}\gt^{\mu\nu}\right]\nn\\
&=&\frac{16\ii\pi\qB}{3(2\pi)^2m}\int_{-\infty}^{+\infty}\frac{dy}{2\pi}\int_{-\infty}^{+\infty}\frac{dx}{2\pi}\left[\frac{1}{(x^2+y^2+1)^2\sqrt{y^2+1}}\gp^{\mu\nu}-\frac{x^2+y^2+1}{(x^2+y^2+1)^2\sqrt{y^2+1}}\gt^{\mu\nu}\right]
\eea

We have the following integrals:
\begin{subequations}
\bea
\int_{-\infty}^{+\infty}\frac{dy}{2\pi}\frac{1}{\sqrt{y^2+1}}\int_{-\infty}^{+\infty}\frac{dx}{2\pi}\frac{1}{(x^2+\alpha_+^2)^2}=\frac{1}{4}\int_{-\infty}^{+\infty}\frac{dy}{2\pi}\frac{1}{(y^2+1)^2}=\frac{1}{16},
\eea
and
\bea
\int_{-\infty}^{+\infty}\frac{dy}{2\pi}\frac{1}{\sqrt{y^2+1}}\int_{-\infty}^{+\infty}\frac{dx}{2\pi}\frac{x^2+y^2+1}{(x^2+\alpha_+^2)^2}=\frac{1}{2}\int_{-\infty}^{+\infty}\frac{dy}{2\pi}\frac{1}{y^2+1}=\frac{1}{4}.
\eea
\end{subequations}

Therefore:
\bea
\lim_{\substack{p_0\to0\\ \mathbf{p}\to0}}T_{3}^{\mu\nu}=\frac{\ii\qB}{3\pi m}\left(\frac{1}{4}\gp^{\mu\nu}-\gt^{\mu\nu}\right).
\eea

\end{widetext}

\section{Projections onto the standard tensor basis}
The tensors $T_i^{\mu\nu}$ can be written as:
\begin{subequations}
    \bea
T_1^{\mu\nu}=\mathcal{I}_1(p)P_\parallel^{\mu\nu}+\mathcal{J}_1(p)g_\parallel^{\mu\nu},
\eea
\bea
T_2^{\mu\nu}=\mathcal{I}_2(p)\left(g_\parallel^{\mu\nu}+2b^\mu b^\nu\right),
\eea
    \bea
T_3^{\mu\nu}=\mathcal{I}_3(p)P_\parallel^{\mu\nu}+\mathcal{J}_3(p)g_\parallel^{\mu\nu}+\mathcal{K}_3(p)g_\perp^{\mu\nu},
\eea
\end{subequations}
where we defined the 4-vector $b^\mu=(0,0,0,1)$, the tensor
\bea
P_\parallel^{\mu\nu}=g_\parallel^{\mu\nu}-\frac{\pp^\mu\pp^\nu}{\pp^2},
\eea
and the integrals
\begin{widetext}
\begin{subequations}
    \bea
\mathcal{I}_1(p)&=&\frac{64\ii\pi\qB}{(2\pi)^2}\pp^2 \exp\left(-\frac{2\pt^2}{3\qB}\right)\int_0^\infty\frac{\lambda^2 d\lambda}{(1+\lambda)^5}\int\frac{d^2\ell_\parallel}{(2\pi)^2}\frac{(1-\lambda)\lp^2+\frac{\lambda^2}{(1+\lambda)^2}\pp^2+m^2}{\left(\ell_\parallel^2+\frac{\lambda}{(1+\lambda)^2}\pp^2-m^2\right)^3\sqrt{\left(l_0-\frac{\lambda}{1+\lambda}p_0\right)^2-m^2}}
\eea
\bea
\mathcal{J}_1(p)&=&\frac{32\ii\pi\qB}{(2\pi)^2}\pp^2 \exp\left(-\frac{2\pt^2}{3\qB}\right)\int_0^\infty\frac{\lambda^2 d\lambda}{(1+\lambda)^5}\int\frac{d^2\ell_\parallel}{(2\pi)^2}\frac{1}{\left(\ell_\parallel^2+\frac{\lambda}{(1+\lambda)^2}\pp^2-m^2\right)^3\sqrt{\left(l_0-\frac{\lambda}{1+\lambda}p_0\right)^2-m^2}}\nn\\
&\times&\left[\left(\lp^2+\frac{\lambda^2}{(1+\lambda)^2}\pp^2+m^2\right)\left(m^2-\frac{\lambda}{(1+\lambda)^2}\pp^2\right)-\frac{\lambda(1-\lambda)}{(1+\lambda)^2}\pp^2\lp^2\right],
\eea
    \bea
    \mathcal{I}_2(p)&=&\frac{16\ii\pi\qB}{(2\pi)^2}\exp\left(-\frac{2\pt^2}{3\qB}\right)\int_0^\infty\frac{d\lambda}{(1+\lambda)^2}\int\frac{d^2\lp}{(2\pi)^2}\frac{-(\ell^3)^2+\frac{\lambda}{(1+\lambda)^2}p_3^2}{\left(\lp^2+\frac{\lambda}{(1+\lambda)^2}\pp^2-m^2\right)^2\sqrt{\left(l_0-\frac{\lambda}{1+\lambda}p_0\right)^2-m^2}},
    \eea
        \bea
\mathcal{I}_3(p)&=&\frac{16\ii\pi\qB}{3(2\pi)^2}\pp^2\exp\left(-\frac{2\pt^2}{3\qB}\right)\int_0^{\infty}\frac{\lambda d\lambda}{(1+\lambda)^4}\int\frac{d^2\lp}{(2\pi)^2}\frac{1}{\sqrt{\left(l_0^2+\frac{\lambda}{1+\lambda}p_0\right)^2-m^2}\left(\lp^2+\frac{\lambda}{(1+\lambda)^2}\pp^2-m^2\right)^2},
\eea
    \bea
\mathcal{J}_3(p)&=&\frac{16\ii\pi\qB}{3(2\pi)^2}\exp\left(-\frac{2\pt^2}{3\qB}\right)\int_0^{\infty}\frac{d\lambda}{(1+\lambda)^2}\int\frac{d^2\lp}{(2\pi)^2}\frac{m^2-\frac{\lambda}{(1+\lambda)^2}\pp^2}{\sqrt{\left(l_0^2+\frac{\lambda}{1+\lambda}p_0\right)^2-m^2}\left(\lp^2+\frac{\lambda}{(1+\lambda)^2}\pp^2-m^2\right)^2},
\eea
    \bea
\mathcal{K}_3(p)&=&-\frac{16\ii\pi\qB}{3(2\pi)^2}\exp\left(-\frac{2\pt^2}{3\qB}\right)\int_0^{\infty}\frac{d\lambda}{(1+\lambda)^2}\int\frac{d^2\lp}{(2\pi)^2}\frac{m^2-\lp^2+\frac{\lambda}{(1+\lambda)^2}\pp^2}{\sqrt{\left(l_0^2+\frac{\lambda}{1+\lambda}p_0\right)^2-m^2}\left(\lp^2+\frac{\lambda}{(1+\lambda)^2}\pp^2-m^2\right)^2}.
\eea
\end{subequations}
\end{widetext}

Let us assume that the full polarization tensor is written in the basis $\left\{P_\parallel^{\mu\nu},P_\perp^{\mu\nu},P_0^{\mu\nu}\right\}$, where
\bea
P_\parallel^{\mu\nu}&\equiv& g_\parallel^{\mu\nu}-\frac{\pp^\mu\pp^\nu}{\pp^2}\nn\\
P_\perp^{\mu\nu}&\equiv& g_\perp^{\mu\nu}+\frac{p_\perp^\mu p_\perp^\nu}{\pt^2},\nn\\
P_0^{\mu\nu}&\equiv& g^{\mu\nu}-\frac{p^\mu p^\nu}{p^2}-P_\parallel^{\mu\nu}-P_\perp^{\mu\nu},
\eea
so that
\bea
\ii\Pi_\Delta^{\mu\nu}=\Pi_\parallel P_\parallel^{\mu\nu}+\Pi_\perp P_\perp^{\mu\nu}+\Pi_0 P_0^{\mu\nu}.
\eea

Hence, given that
\bea
\ii\Pi_\Delta^{\mu\nu}&=&\ii\Pi_0^{\mu\nu}+\ii\frac{q^2\qB\Delta}{4\pi}\sum_{i=1}^{3}T_i^{\mu\nu}\nn\\
&=&\ii\frac{q^2\qB}{4\pi^2}\Bigg\{\left[e^{-\frac{\pt^2}{2\qB}}\pp^2\mathcal{I}_0+\Delta\left(\mathcal{I}_1+\mathcal{I}_3\right)\right]P_\parallel^{\mu\nu}.\nn\\
&+&\Delta\left(\mathcal{J}_1+\mathcal{I}_2+\mathcal{J}_3\right)\gp^{\mu\nu}+\Delta\mathcal{K}_3\gt^{\mu\nu}+2\Delta\mathcal{I}_2 b^\mu b^\nu\Bigg\},\nn\\
\eea
it is straightforward to show that:
\begin{subequations}
\bea
\Pi_\parallel&=&\ii\frac{q^2\qB}{4\pi}\Bigg[e^{-\frac{\pt^2}{2\qB}}\pp^2\mathcal{I}_0+\Delta\left(\mathcal{I}_1+\mathcal{I}_3\right)-\frac{2\Delta p_0^2}{\pp^2}\mathcal{I}_2\nn\\
&+&\Delta\left(\mathcal{J}_1+\mathcal{I}_2+\mathcal{J}_3\right)\Bigg],\nn\\
\eea
\bea
\Pi_\perp=\ii\frac{q^2\qB}{4\pi}\Delta\mathcal{K}_3,
\eea
and
\bea
\Pi_0&=&\ii\frac{q^2\qB}{4\pi}\Bigg[-\frac{\pt^2}{p^2}\Delta\left(\mathcal{J}_1+\mathcal{I}_2+\mathcal{J}_3\right)+\frac{\pp^2}{p^2}\Delta\mathcal{K}_3\nn\\
&+&2\Delta\left(\frac{p_0^2}{\pp^2}-\frac{p_3^2}{p^2}-1\right)\mathcal{I}_2\Bigg],
\eea
\end{subequations}
provided by the fact that the basis is orthonormal and
\begin{subequations}
    \bea
    P_\parallel^{\mu\nu}g_{\mu\nu}^\parallel=1
    \eea
    \bea
    P_\parallel^{\mu\nu}b_\mu b_\nu=-\frac{p_0^2}{\pp^2}
    \eea
    \bea
    P_\perp^{\mu\nu}g_{\mu\nu}^\perp=1
    \eea
    \bea
    P_0^{\mu\nu}g_{\mu\nu}^\parallel=-\frac{\pt^2}{p^2}
    \eea
    \bea
    P_0^{\mu\nu}b_\mu b_\nu=\frac{p_0^2}{\pp^2}-\frac{p_3^2}{p^2}-1,
    \eea
    and
    \bea
    P_0^{\mu\nu}g_{\mu\nu}^\perp=\frac{\pp^2}{p^2}.
    \eea
\end{subequations}


\begin{thebibliography}{28}%
\makeatletter
\providecommand \@ifxundefined [1]{%
 \@ifx{#1\undefined}
}%
\providecommand \@ifnum [1]{%
 \ifnum #1\expandafter \@firstoftwo
 \else \expandafter \@secondoftwo
 \fi
}%
\providecommand \@ifx [1]{%
 \ifx #1\expandafter \@firstoftwo
 \else \expandafter \@secondoftwo
 \fi
}%
\providecommand \natexlab [1]{#1}%
\providecommand \enquote  [1]{``#1''}%
\providecommand \bibnamefont  [1]{#1}%
\providecommand \bibfnamefont [1]{#1}%
\providecommand \citenamefont [1]{#1}%
\providecommand \href@noop [0]{\@secondoftwo}%
\providecommand \href [0]{\begingroup \@sanitize@url \@href}%
\providecommand \@href[1]{\@@startlink{#1}\@@href}%
\providecommand \@@href[1]{\endgroup#1\@@endlink}%
\providecommand \@sanitize@url [0]{\catcode `\\12\catcode `\$12\catcode
  `\&12\catcode `\#12\catcode `\^12\catcode `\_12\catcode `\%12\relax}%
\providecommand \@@startlink[1]{}%
\providecommand \@@endlink[0]{}%
\providecommand \url  [0]{\begingroup\@sanitize@url \@url }%
\providecommand \@url [1]{\endgroup\@href {#1}{\urlprefix }}%
\providecommand \urlprefix  [0]{URL }%
\providecommand \Eprint [0]{\href }%
\providecommand \doibase [0]{http://dx.doi.org/}%
\providecommand \selectlanguage [0]{\@gobble}%
\providecommand \bibinfo  [0]{\@secondoftwo}%
\providecommand \bibfield  [0]{\@secondoftwo}%
\providecommand \translation [1]{[#1]}%
\providecommand \BibitemOpen [0]{}%
\providecommand \bibitemStop [0]{}%
\providecommand \bibitemNoStop [0]{.\EOS\space}%
\providecommand \EOS [0]{\spacefactor3000\relax}%
\providecommand \BibitemShut  [1]{\csname bibitem#1\endcsname}%
\let\auto@bib@innerbib\@empty
\bibitem [{\citenamefont {David}(2020)}]{David_2020}%
  \BibitemOpen
  \bibfield  {author} {\bibinfo {author} {\bibfnamefont {Gabor}\ \bibnamefont
  {David}},\ }\bibfield  {title} {\enquote {\bibinfo {title} {Direct real
  photons in relativistic heavy ion collisions},}\ }\href {\doibase
  10.1088/1361-6633/ab6f57} {\bibfield  {journal} {\bibinfo  {journal} {Reports
  on Progress in Physics}\ }\textbf {\bibinfo {volume} {83}},\ \bibinfo {pages}
  {046301} (\bibinfo {year} {2020})}\BibitemShut {NoStop}%
\bibitem [{\citenamefont {Adare}\ \emph {et~al.}(2015)\citenamefont {Adare}
  \emph {et~al.}}]{PhysRevC.91.064904}%
  \BibitemOpen
  \bibfield  {author} {\bibinfo {author} {\bibfnamefont {A.}~\bibnamefont
  {Adare}} \emph {et~al.} (\bibinfo {collaboration} {PHENIX Collaboration}),\
  }\bibfield  {title} {\enquote {\bibinfo {title} {Centrality dependence of
  low-momentum direct-photon production in $\mathrm{Au}+\mathrm{Au}$ collisions
  at $\sqrt{{s}_{\mathit{N}N}}=200 \mathrm{GeV}$},}\ }\href {\doibase
  10.1103/PhysRevC.91.064904} {\bibfield  {journal} {\bibinfo  {journal} {Phys.
  Rev. C}\ }\textbf {\bibinfo {volume} {91}},\ \bibinfo {pages} {064904}
  (\bibinfo {year} {2015})}\BibitemShut {NoStop}%
\bibitem [{\citenamefont {Adare}\ \emph {et~al.}(2012)\citenamefont {Adare}
  \emph {et~al.}}]{PhysRevLett.109.122302}%
  \BibitemOpen
  \bibfield  {author} {\bibinfo {author} {\bibfnamefont {A.}~\bibnamefont
  {Adare}} \emph {et~al.} (\bibinfo {collaboration} {PHENIX Collaboration}),\
  }\bibfield  {title} {\enquote {\bibinfo {title} {Observation of direct-photon
  collective flow in $\mathrm{Au}+\mathrm{Au}$ collisions at
  $\sqrt{{s}_{NN}}=200\text{ }\text{ }\mathrm{GeV}$},}\ }\href {\doibase
  10.1103/PhysRevLett.109.122302} {\bibfield  {journal} {\bibinfo  {journal}
  {Phys. Rev. Lett.}\ }\textbf {\bibinfo {volume} {109}},\ \bibinfo {pages}
  {122302} (\bibinfo {year} {2012})}\BibitemShut {NoStop}%
\bibitem [{\citenamefont {Acharya}\ \emph {et~al.}(2019)\citenamefont {Acharya}
  \emph {et~al.}}]{2019308}%
  \BibitemOpen
  \bibfield  {author} {\bibinfo {author} {\bibfnamefont {S.}~\bibnamefont
  {Acharya}} \emph {et~al.},\ }\bibfield  {title} {\enquote {\bibinfo {title}
  {Direct photon elliptic flow in pb–pb collisions at snn=2.76 tev},}\ }\href
  {\doibase https://doi.org/10.1016/j.physletb.2018.11.039} {\bibfield
  {journal} {\bibinfo  {journal} {Physics Letters B}\ }\textbf {\bibinfo
  {volume} {789}},\ \bibinfo {pages} {308--322} (\bibinfo {year}
  {2019})}\BibitemShut {NoStop}%
\bibitem [{\citenamefont {Adare}\ \emph {et~al.}(2016)\citenamefont {Adare}
  \emph {et~al.}}]{PhysRevC.94.064901}%
  \BibitemOpen
  \bibfield  {author} {\bibinfo {author} {\bibfnamefont {A.}~\bibnamefont
  {Adare}} \emph {et~al.} (\bibinfo {collaboration} {PHENIX Collaboration}),\
  }\bibfield  {title} {\enquote {\bibinfo {title} {Azimuthally anisotropic
  emission of low-momentum direct photons in au + au collisions at
  $\sqrt{{s}_{NN}}=200$ gev},}\ }\href {\doibase 10.1103/PhysRevC.94.064901}
  {\bibfield  {journal} {\bibinfo  {journal} {Phys. Rev. C}\ }\textbf {\bibinfo
  {volume} {94}},\ \bibinfo {pages} {064901} (\bibinfo {year}
  {2016})}\BibitemShut {NoStop}%
\bibitem [{\citenamefont {Ghiglieri}(2014)}]{GHIGLIERI2014326}%
  \BibitemOpen
  \bibfield  {author} {\bibinfo {author} {\bibfnamefont {Jacopo}\ \bibnamefont
  {Ghiglieri}},\ }\bibfield  {title} {\enquote {\bibinfo {title}
  {Next-to-leading order thermal photon production in a weakly-coupled
  plasma},}\ }\href {\doibase https://doi.org/10.1016/j.nuclphysa.2014.09.048}
  {\bibfield  {journal} {\bibinfo  {journal} {Nuclear Physics A}\ }\textbf
  {\bibinfo {volume} {932}},\ \bibinfo {pages} {326--333} (\bibinfo {year}
  {2014})},\ \bibinfo {note} {hard Probes 2013}\BibitemShut {NoStop}%
\bibitem [{\citenamefont {Paquet}\ \emph
  {et~al.}(2016{\natexlab{a}})\citenamefont {Paquet}, \citenamefont {Shen},
  \citenamefont {Denicol}, \citenamefont {Luzum}, \citenamefont {Schenke},
  \citenamefont {Jeon},\ and\ \citenamefont {Gale}}]{PAQUET2016409}%
  \BibitemOpen
  \bibfield  {author} {\bibinfo {author} {\bibfnamefont {Jean-François}\
  \bibnamefont {Paquet}}, \bibinfo {author} {\bibfnamefont {Chun}\ \bibnamefont
  {Shen}}, \bibinfo {author} {\bibfnamefont {Gabriel}\ \bibnamefont {Denicol}},
  \bibinfo {author} {\bibfnamefont {Matthew}\ \bibnamefont {Luzum}}, \bibinfo
  {author} {\bibfnamefont {Björn}\ \bibnamefont {Schenke}}, \bibinfo {author}
  {\bibfnamefont {Sangyong}\ \bibnamefont {Jeon}}, \ and\ \bibinfo {author}
  {\bibfnamefont {Charles}\ \bibnamefont {Gale}},\ }\bibfield  {title}
  {\enquote {\bibinfo {title} {Thermal and prompt photons at rhic and the
  lhc},}\ }\href {\doibase https://doi.org/10.1016/j.nuclphysa.2016.01.068}
  {\bibfield  {journal} {\bibinfo  {journal} {Nuclear Physics A}\ }\textbf
  {\bibinfo {volume} {956}},\ \bibinfo {pages} {409--412} (\bibinfo {year}
  {2016}{\natexlab{a}})},\ \bibinfo {note} {the XXV International Conference on
  Ultrarelativistic Nucleus-Nucleus Collisions: Quark Matter 2015}\BibitemShut
  {NoStop}%
\bibitem [{\citenamefont {Paquet}\ \emph
  {et~al.}(2016{\natexlab{b}})\citenamefont {Paquet}, \citenamefont {Shen},
  \citenamefont {Denicol}, \citenamefont {Luzum}, \citenamefont {Schenke},
  \citenamefont {Jeon},\ and\ \citenamefont {Gale}}]{PhysRevC.93.044906}%
  \BibitemOpen
  \bibfield  {author} {\bibinfo {author} {\bibfnamefont {Jean-Fran\ifmmode
  \mbox{\c{c}}\else~\c{c}\fi{}ois}\ \bibnamefont {Paquet}}, \bibinfo {author}
  {\bibfnamefont {Chun}\ \bibnamefont {Shen}}, \bibinfo {author} {\bibfnamefont
  {Gabriel~S.}\ \bibnamefont {Denicol}}, \bibinfo {author} {\bibfnamefont
  {Matthew}\ \bibnamefont {Luzum}}, \bibinfo {author} {\bibfnamefont {Bj\"orn}\
  \bibnamefont {Schenke}}, \bibinfo {author} {\bibfnamefont {Sangyong}\
  \bibnamefont {Jeon}}, \ and\ \bibinfo {author} {\bibfnamefont {Charles}\
  \bibnamefont {Gale}},\ }\bibfield  {title} {\enquote {\bibinfo {title}
  {Production of photons in relativistic heavy-ion collisions},}\ }\href
  {\doibase 10.1103/PhysRevC.93.044906} {\bibfield  {journal} {\bibinfo
  {journal} {Phys. Rev. C}\ }\textbf {\bibinfo {volume} {93}},\ \bibinfo
  {pages} {044906} (\bibinfo {year} {2016}{\natexlab{b}})}\BibitemShut
  {NoStop}%
\bibitem [{\citenamefont {Ayala}\ \emph {et~al.}(2022)\citenamefont {Ayala},
  \citenamefont {Casta\~no Yepes}, \citenamefont {Hern\'andez}, \citenamefont
  {Mizher}, \citenamefont {Tejeda-Yeomans},\ and\ \citenamefont
  {Zamora}}]{PhysRevC.106.064905}%
  \BibitemOpen
  \bibfield  {author} {\bibinfo {author} {\bibfnamefont {Alejandro}\
  \bibnamefont {Ayala}}, \bibinfo {author} {\bibfnamefont {Jorge~David}\
  \bibnamefont {Casta\~no Yepes}}, \bibinfo {author} {\bibfnamefont {L.~A.}\
  \bibnamefont {Hern\'andez}}, \bibinfo {author} {\bibfnamefont {Ana~Julia}\
  \bibnamefont {Mizher}}, \bibinfo {author} {\bibfnamefont {Mar\'{\i}a~Elena}\
  \bibnamefont {Tejeda-Yeomans}}, \ and\ \bibinfo {author} {\bibfnamefont
  {R.}~\bibnamefont {Zamora}},\ }\bibfield  {title} {\enquote {\bibinfo {title}
  {Anisotropic photon emission from gluon fusion and splitting in a strong
  magnetic background: The two-gluon one-photon vertex},}\ }\href {\doibase
  10.1103/PhysRevC.106.064905} {\bibfield  {journal} {\bibinfo  {journal}
  {Phys. Rev. C}\ }\textbf {\bibinfo {volume} {106}},\ \bibinfo {pages}
  {064905} (\bibinfo {year} {2022})}\BibitemShut {NoStop}%
\bibitem [{\citenamefont {Ayala}\ \emph
  {et~al.}(2020{\natexlab{a}})\citenamefont {Ayala}, \citenamefont
  {Castaño-Yepes}, \citenamefont {Dominguez~Jimenez}, \citenamefont {Salinas
  San~Martín},\ and\ \citenamefont {Tejeda-Yeomans}}]{Ayala2020}%
  \BibitemOpen
  \bibfield  {author} {\bibinfo {author} {\bibfnamefont {Alejandro}\
  \bibnamefont {Ayala}}, \bibinfo {author} {\bibfnamefont {Jorge~David}\
  \bibnamefont {Castaño-Yepes}}, \bibinfo {author} {\bibfnamefont {Isabel}\
  \bibnamefont {Dominguez~Jimenez}}, \bibinfo {author} {\bibfnamefont {Jordi}\
  \bibnamefont {Salinas San~Martín}}, \ and\ \bibinfo {author} {\bibfnamefont
  {María~Elena}\ \bibnamefont {Tejeda-Yeomans}},\ }\bibfield  {title}
  {\enquote {\bibinfo {title} {Centrality dependence of photon yield and
  elliptic flow from gluon fusion and splitting induced by magnetic fields in
  relativistic heavy-ion collisions},}\ }\href {\doibase
  10.1140/epja/s10050-020-00060-9} {\bibfield  {journal} {\bibinfo  {journal}
  {Eur. Phys. J. A}\ }\textbf {\bibinfo {volume} {56}},\ \bibinfo {pages} {53}
  (\bibinfo {year} {2020}{\natexlab{a}})}\BibitemShut {NoStop}%
\bibitem [{\citenamefont {Ayala}\ \emph {et~al.}(2017)\citenamefont {Ayala},
  \citenamefont {Castano-Yepes}, \citenamefont {Dominguez}, \citenamefont
  {Hernandez}, \citenamefont {Hernandez-Ortiz},\ and\ \citenamefont
  {Tejeda-Yeomans}}]{Ayala:2017vex}%
  \BibitemOpen
  \bibfield  {author} {\bibinfo {author} {\bibfnamefont {Alejandro}\
  \bibnamefont {Ayala}}, \bibinfo {author} {\bibfnamefont {Jorge~David}\
  \bibnamefont {Castano-Yepes}}, \bibinfo {author} {\bibfnamefont {Cesareo~A.}\
  \bibnamefont {Dominguez}}, \bibinfo {author} {\bibfnamefont {Luis~A.}\
  \bibnamefont {Hernandez}}, \bibinfo {author} {\bibfnamefont {Saul}\
  \bibnamefont {Hernandez-Ortiz}}, \ and\ \bibinfo {author} {\bibfnamefont
  {Maria~Elena}\ \bibnamefont {Tejeda-Yeomans}},\ }\bibfield  {title} {\enquote
  {\bibinfo {title} {{Prompt photon yield and elliptic flow from gluon fusion
  induced by magnetic fields in relativistic heavy-ion collisions}},}\ }\href
  {\doibase 10.1103/PhysRevD.96.014023} {\bibfield  {journal} {\bibinfo
  {journal} {Phys. Rev. D}\ }\textbf {\bibinfo {volume} {96}},\ \bibinfo
  {pages} {014023} (\bibinfo {year} {2017})},\ \bibinfo {note} {[Erratum:
  Phys.Rev.D 96, 119901 (2017)]}\BibitemShut {NoStop}%
\bibitem [{\citenamefont {Jia}\ \emph {et~al.}(2023)\citenamefont {Jia},
  \citenamefont {Li},\ and\ \citenamefont {Hou}}]{JIA2023138239}%
  \BibitemOpen
  \bibfield  {author} {\bibinfo {author} {\bibfnamefont {Moran}\ \bibnamefont
  {Jia}}, \bibinfo {author} {\bibfnamefont {Huixia}\ \bibnamefont {Li}}, \ and\
  \bibinfo {author} {\bibfnamefont {Defu}\ \bibnamefont {Hou}},\ }\bibfield
  {title} {\enquote {\bibinfo {title} {The photon production and collective
  flows from magnetic induced gluon fusion and splitting in early stage of high
  energy nuclear collision},}\ }\href {\doibase
  https://doi.org/10.1016/j.physletb.2023.138239} {\bibfield  {journal}
  {\bibinfo  {journal} {Physics Letters B}\ }\textbf {\bibinfo {volume}
  {846}},\ \bibinfo {pages} {138239} (\bibinfo {year} {2023})}\BibitemShut
  {NoStop}%
\bibitem [{\citenamefont {Lappi}\ and\ \citenamefont
  {McLerran}(2006)}]{LAPPI2006200}%
  \BibitemOpen
  \bibfield  {author} {\bibinfo {author} {\bibfnamefont {T.}~\bibnamefont
  {Lappi}}\ and\ \bibinfo {author} {\bibfnamefont {L.}~\bibnamefont
  {McLerran}},\ }\bibfield  {title} {\enquote {\bibinfo {title} {Some features
  of the glasma},}\ }\href {\doibase
  https://doi.org/10.1016/j.nuclphysa.2006.04.001} {\bibfield  {journal}
  {\bibinfo  {journal} {Nuclear Physics A}\ }\textbf {\bibinfo {volume}
  {772}},\ \bibinfo {pages} {200--212} (\bibinfo {year} {2006})}\BibitemShut
  {NoStop}%
\bibitem [{\citenamefont {McLerran}\ and\ \citenamefont
  {Schenke}(2014)}]{MCLERRAN201471}%
  \BibitemOpen
  \bibfield  {author} {\bibinfo {author} {\bibfnamefont {Larry}\ \bibnamefont
  {McLerran}}\ and\ \bibinfo {author} {\bibfnamefont {Björn}\ \bibnamefont
  {Schenke}},\ }\bibfield  {title} {\enquote {\bibinfo {title} {{The Glasma,
  photons and the implications of anisotropy}},}\ }\href {\doibase
  https://doi.org/10.1016/j.nuclphysa.2014.06.004} {\bibfield  {journal}
  {\bibinfo  {journal} {Nuclear Physics A}\ }\textbf {\bibinfo {volume}
  {929}},\ \bibinfo {pages} {71--82} (\bibinfo {year} {2014})}\BibitemShut
  {NoStop}%
\bibitem [{\citenamefont {Harland-Lang}\ \emph {et~al.}(2015)\citenamefont
  {Harland-Lang}, \citenamefont {Martin}, \citenamefont {Motylinski},\ and\
  \citenamefont {Thorne}}]{Harland-Lang2015}%
  \BibitemOpen
  \bibfield  {author} {\bibinfo {author} {\bibfnamefont {L.~A.}\ \bibnamefont
  {Harland-Lang}}, \bibinfo {author} {\bibfnamefont {A.~D.}\ \bibnamefont
  {Martin}}, \bibinfo {author} {\bibfnamefont {P.}~\bibnamefont {Motylinski}},
  \ and\ \bibinfo {author} {\bibfnamefont {R.~S.}\ \bibnamefont {Thorne}},\
  }\bibfield  {title} {\enquote {\bibinfo {title} {{Parton distributions in the
  LHC era: MMHT 2014 PDFs}},}\ }\href {\doibase 10.1140/epjc/s10052-015-3397-6}
  {\bibfield  {journal} {\bibinfo  {journal} {Eur. Phys. J. C}\ }\textbf
  {\bibinfo {volume} {75}},\ \bibinfo {pages} {204} (\bibinfo {year}
  {2015})}\BibitemShut {NoStop}%
\bibitem [{\citenamefont {Carrington}\ \emph {et~al.}(2022)\citenamefont
  {Carrington}, \citenamefont {Czajka},\ and\ \citenamefont
  {Mr\'owczy\ifmmode~\acute{n}\else \'{n}\fi{}ski}}]{PhysRevC.106.034904}%
  \BibitemOpen
  \bibfield  {author} {\bibinfo {author} {\bibfnamefont {Margaret~E.}\
  \bibnamefont {Carrington}}, \bibinfo {author} {\bibfnamefont {Alina}\
  \bibnamefont {Czajka}}, \ and\ \bibinfo {author} {\bibfnamefont
  {Stanis\l{}aw}\ \bibnamefont {Mr\'owczy\ifmmode~\acute{n}\else
  \'{n}\fi{}ski}},\ }\bibfield  {title} {\enquote {\bibinfo {title} {Physical
  characteristics of glasma from the earliest stage of relativistic heavy ion
  collisions},}\ }\href {\doibase 10.1103/PhysRevC.106.034904} {\bibfield
  {journal} {\bibinfo  {journal} {Phys. Rev. C}\ }\textbf {\bibinfo {volume}
  {106}},\ \bibinfo {pages} {034904} (\bibinfo {year} {2022})}\BibitemShut
  {NoStop}%
\bibitem [{\citenamefont {Adler}\ \emph {et~al.}(1970)\citenamefont {Adler},
  \citenamefont {Bahcall}, \citenamefont {Callan},\ and\ \citenamefont
  {Rosenbluth}}]{PhysRevLett.25.1061}%
  \BibitemOpen
  \bibfield  {author} {\bibinfo {author} {\bibfnamefont {S.~L.}\ \bibnamefont
  {Adler}}, \bibinfo {author} {\bibfnamefont {J.~N.}\ \bibnamefont {Bahcall}},
  \bibinfo {author} {\bibfnamefont {C.~G.}\ \bibnamefont {Callan}}, \ and\
  \bibinfo {author} {\bibfnamefont {M.~N.}\ \bibnamefont {Rosenbluth}},\
  }\bibfield  {title} {\enquote {\bibinfo {title} {Photon splitting in a strong
  magnetic field},}\ }\href {\doibase 10.1103/PhysRevLett.25.1061} {\bibfield
  {journal} {\bibinfo  {journal} {Phys. Rev. Lett.}\ }\textbf {\bibinfo
  {volume} {25}},\ \bibinfo {pages} {1061--1065} (\bibinfo {year}
  {1970})}\BibitemShut {NoStop}%
\bibitem [{\citenamefont {Nieves}\ \emph {et~al.}(1983)\citenamefont {Nieves},
  \citenamefont {Pal},\ and\ \citenamefont {Unger}}]{PhysRevD.28.908}%
  \BibitemOpen
  \bibfield  {author} {\bibinfo {author} {\bibfnamefont {Jos\'e~F.}\
  \bibnamefont {Nieves}}, \bibinfo {author} {\bibfnamefont {Palash~B.}\
  \bibnamefont {Pal}}, \ and\ \bibinfo {author} {\bibfnamefont {David~G.}\
  \bibnamefont {Unger}},\ }\bibfield  {title} {\enquote {\bibinfo {title}
  {Photon mass in a background of thermal particles},}\ }\href {\doibase
  10.1103/PhysRevD.28.908} {\bibfield  {journal} {\bibinfo  {journal} {Phys.
  Rev. D}\ }\textbf {\bibinfo {volume} {28}},\ \bibinfo {pages} {908--914}
  (\bibinfo {year} {1983})}\BibitemShut {NoStop}%
\bibitem [{\citenamefont {Le~Bellac}(1996)}]{lebellac}%
  \BibitemOpen
  \bibfield  {author} {\bibinfo {author} {\bibfnamefont {Michel}\ \bibnamefont
  {Le~Bellac}},\ }\href@noop {} {\emph {\bibinfo {title} {Thermal Field
  Theory}}}\ (\bibinfo  {publisher} {Cambridge University Press},\ \bibinfo
  {address} {Cambridge, UK},\ \bibinfo {year} {1996})\BibitemShut {NoStop}%
\bibitem [{\citenamefont {Voronyuk}\ \emph {et~al.}(2011)\citenamefont
  {Voronyuk}, \citenamefont {Toneev}, \citenamefont {Cassing}, \citenamefont
  {Bratkovskaya}, \citenamefont {Konchakovski},\ and\ \citenamefont
  {Voloshin}}]{PhysRevC.83.054911}%
  \BibitemOpen
  \bibfield  {author} {\bibinfo {author} {\bibfnamefont {V.}~\bibnamefont
  {Voronyuk}}, \bibinfo {author} {\bibfnamefont {V.~D.}\ \bibnamefont
  {Toneev}}, \bibinfo {author} {\bibfnamefont {W.}~\bibnamefont {Cassing}},
  \bibinfo {author} {\bibfnamefont {E.~L.}\ \bibnamefont {Bratkovskaya}},
  \bibinfo {author} {\bibfnamefont {V.~P.}\ \bibnamefont {Konchakovski}}, \
  and\ \bibinfo {author} {\bibfnamefont {S.~A.}\ \bibnamefont {Voloshin}},\
  }\bibfield  {title} {\enquote {\bibinfo {title} {{Electromagnetic field
  evolution in relativistic heavy-ion collisions}},}\ }\href {\doibase
  10.1103/PhysRevC.83.054911} {\bibfield  {journal} {\bibinfo  {journal} {Phys.
  Rev. C}\ }\textbf {\bibinfo {volume} {83}},\ \bibinfo {pages} {054911}
  (\bibinfo {year} {2011})}\BibitemShut {NoStop}%
\bibitem [{\citenamefont {Adam}\ \emph {et~al.}(2021)\citenamefont {Adam} \emph
  {et~al.}}]{PhysRevLett.127.052302}%
  \BibitemOpen
  \bibfield  {author} {\bibinfo {author} {\bibfnamefont {J.}~\bibnamefont
  {Adam}} \emph {et~al.} (\bibinfo {collaboration} {STAR Collaboration}),\
  }\bibfield  {title} {\enquote {\bibinfo {title} {Measurement of
  ${e}^{+}{e}^{\ensuremath{-}}$ momentum and angular distributions from
  linearly polarized photon collisions},}\ }\href {\doibase
  10.1103/PhysRevLett.127.052302} {\bibfield  {journal} {\bibinfo  {journal}
  {Phys. Rev. Lett.}\ }\textbf {\bibinfo {volume} {127}},\ \bibinfo {pages}
  {052302} (\bibinfo {year} {2021})}\BibitemShut {NoStop}%
\bibitem [{\citenamefont {Wang}\ \emph {et~al.}(2022)\citenamefont {Wang},
  \citenamefont {Zhao}, \citenamefont {Greiner}, \citenamefont {Xu},\ and\
  \citenamefont {Zhuang}}]{PhysRevC.105.L041901}%
  \BibitemOpen
  \bibfield  {author} {\bibinfo {author} {\bibfnamefont {Zeyan}\ \bibnamefont
  {Wang}}, \bibinfo {author} {\bibfnamefont {Jiaxing}\ \bibnamefont {Zhao}},
  \bibinfo {author} {\bibfnamefont {Carsten}\ \bibnamefont {Greiner}}, \bibinfo
  {author} {\bibfnamefont {Zhe}\ \bibnamefont {Xu}}, \ and\ \bibinfo {author}
  {\bibfnamefont {Pengfei}\ \bibnamefont {Zhuang}},\ }\bibfield  {title}
  {\enquote {\bibinfo {title} {{Incomplete electromagnetic response of hot QCD
  matter}},}\ }\href {\doibase 10.1103/PhysRevC.105.L041901} {\bibfield
  {journal} {\bibinfo  {journal} {Phys. Rev. C}\ }\textbf {\bibinfo {volume}
  {105}},\ \bibinfo {pages} {L041901} (\bibinfo {year} {2022})}\BibitemShut
  {NoStop}%
\bibitem [{\citenamefont {Ayala}\ \emph {et~al.}(2021)\citenamefont {Ayala},
  \citenamefont {Castaño-Yepes}, \citenamefont {Hernández}, \citenamefont
  {Martín},\ and\ \citenamefont {Zamora}}]{Ayala2021}%
  \BibitemOpen
  \bibfield  {author} {\bibinfo {author} {\bibfnamefont {Alejandro}\
  \bibnamefont {Ayala}}, \bibinfo {author} {\bibfnamefont {Jorge~David}\
  \bibnamefont {Castaño-Yepes}}, \bibinfo {author} {\bibfnamefont {L.~A.}\
  \bibnamefont {Hernández}}, \bibinfo {author} {\bibfnamefont {Jordi
  Salinas~San}\ \bibnamefont {Martín}}, \ and\ \bibinfo {author}
  {\bibfnamefont {R.}~\bibnamefont {Zamora}},\ }\bibfield  {title} {\enquote
  {\bibinfo {title} {Gluon polarization tensor and dispersion relation in a
  weakly magnetized medium},}\ }\href {\doibase
  10.1140/epja/s10050-021-00429-4} {\bibfield  {journal} {\bibinfo  {journal}
  {Eur. Phys. J. A}\ }\textbf {\bibinfo {volume} {57}},\ \bibinfo {pages} {140}
  (\bibinfo {year} {2021})}\BibitemShut {NoStop}%
\bibitem [{\citenamefont {Ayala}\ \emph
  {et~al.}(2020{\natexlab{b}})\citenamefont {Ayala}, \citenamefont {Casta\~no
  Yepes}, \citenamefont {Loewe},\ and\ \citenamefont
  {Mu\~noz}}]{Ayala_Pol_020}%
  \BibitemOpen
  \bibfield  {author} {\bibinfo {author} {\bibfnamefont {Alejandro}\
  \bibnamefont {Ayala}}, \bibinfo {author} {\bibfnamefont {Jorge~David}\
  \bibnamefont {Casta\~no Yepes}}, \bibinfo {author} {\bibfnamefont
  {M.}~\bibnamefont {Loewe}}, \ and\ \bibinfo {author} {\bibfnamefont
  {Enrique}\ \bibnamefont {Mu\~noz}},\ }\bibfield  {title} {\enquote {\bibinfo
  {title} {{Gluon polarization tensor in a magnetized medium: Analytic approach
  starting from the sum over Landau levels}},}\ }\href {\doibase
  10.1103/PhysRevD.101.036016} {\bibfield  {journal} {\bibinfo  {journal}
  {Phys. Rev. D}\ }\textbf {\bibinfo {volume} {101}},\ \bibinfo {pages}
  {036016} (\bibinfo {year} {2020}{\natexlab{b}})}\BibitemShut {NoStop}%
\bibitem [{\citenamefont {Casta\~no Yepes}\ \emph
  {et~al.}(2023{\natexlab{a}})\citenamefont {Casta\~no Yepes}, \citenamefont
  {Loewe}, \citenamefont {Mu\~noz}, \citenamefont {Rojas},\ and\ \citenamefont
  {Zamora}}]{PhysRevD.107.096014}%
  \BibitemOpen
  \bibfield  {author} {\bibinfo {author} {\bibfnamefont {Jorge~David}\
  \bibnamefont {Casta\~no Yepes}}, \bibinfo {author} {\bibfnamefont {Marcelo}\
  \bibnamefont {Loewe}}, \bibinfo {author} {\bibfnamefont {Enrique}\
  \bibnamefont {Mu\~noz}}, \bibinfo {author} {\bibfnamefont {Juan~Crist\'obal}\
  \bibnamefont {Rojas}}, \ and\ \bibinfo {author} {\bibfnamefont {Renato}\
  \bibnamefont {Zamora}},\ }\bibfield  {title} {\enquote {\bibinfo {title}
  {{QED fermions in a noisy magnetic field background}},}\ }\href {\doibase
  10.1103/PhysRevD.107.096014} {\bibfield  {journal} {\bibinfo  {journal}
  {Phys. Rev. D}\ }\textbf {\bibinfo {volume} {107}},\ \bibinfo {pages}
  {096014} (\bibinfo {year} {2023}{\natexlab{a}})}\BibitemShut {NoStop}%
\bibitem [{\citenamefont {Casta\~no Yepes}\ \emph
  {et~al.}(2023{\natexlab{b}})\citenamefont {Casta\~no Yepes}, \citenamefont
  {Loewe}, \citenamefont {Mu\~noz},\ and\ \citenamefont
  {Rojas}}]{PhysRevD.108.116013}%
  \BibitemOpen
  \bibfield  {author} {\bibinfo {author} {\bibfnamefont {Jorge~David}\
  \bibnamefont {Casta\~no Yepes}}, \bibinfo {author} {\bibfnamefont {Marcelo}\
  \bibnamefont {Loewe}}, \bibinfo {author} {\bibfnamefont {Enrique}\
  \bibnamefont {Mu\~noz}}, \ and\ \bibinfo {author} {\bibfnamefont
  {Juan~Crist\'obal}\ \bibnamefont {Rojas}},\ }\bibfield  {title} {\enquote
  {\bibinfo {title} {Qed fermions in a noisy magnetic field background: The
  effective action approach},}\ }\href {\doibase 10.1103/PhysRevD.108.116013}
  {\bibfield  {journal} {\bibinfo  {journal} {Phys. Rev. D}\ }\textbf {\bibinfo
  {volume} {108}},\ \bibinfo {pages} {116013} (\bibinfo {year}
  {2023}{\natexlab{b}})}\BibitemShut {NoStop}%
\bibitem [{\citenamefont {M{\'e}zard}\ and\ \citenamefont
  {Parisi}(1991)}]{mezard1991replica}%
  \BibitemOpen
  \bibfield  {author} {\bibinfo {author} {\bibfnamefont {Marc}\ \bibnamefont
  {M{\'e}zard}}\ and\ \bibinfo {author} {\bibfnamefont {Giorgio}\ \bibnamefont
  {Parisi}},\ }\bibfield  {title} {\enquote {\bibinfo {title} {Replica field
  theory for random manifolds},}\ }\href {\doibase
  https://doi.org/10.1051/jp1:1991171} {\bibfield  {journal} {\bibinfo
  {journal} {Journal de Physique I}\ }\textbf {\bibinfo {volume} {1}},\
  \bibinfo {pages} {809--836} (\bibinfo {year} {1991})}\BibitemShut {NoStop}%
\bibitem [{\citenamefont {Fukushima}(2011)}]{PhysRevD.83.111501}%
  \BibitemOpen
  \bibfield  {author} {\bibinfo {author} {\bibfnamefont {Kenji}\ \bibnamefont
  {Fukushima}},\ }\bibfield  {title} {\enquote {\bibinfo {title}
  {Magnetic-field induced screening effect and collective excitations},}\
  }\href {\doibase 10.1103/PhysRevD.83.111501} {\bibfield  {journal} {\bibinfo
  {journal} {Phys. Rev. D}\ }\textbf {\bibinfo {volume} {83}},\ \bibinfo
  {pages} {111501} (\bibinfo {year} {2011})}\BibitemShut {NoStop}%
\end{thebibliography}%
\end{document}